\documentclass[aps,prd,amsfonts,amssymb,amsmath,mathrsfs,nofootinbib,reprint,showpacs,fixltx2e,subfig,usegraphicx,twocolumn]{revtex4}
\usepackage{graphicx}
\usepackage[font=small,labelfont=bf,textfont=footnotesize,justification=raggedright,singlelinecheck=true]{caption}
\usepackage{subfig}
\usepackage{aas_macros}
\usepackage{mathrsfs}
\usepackage{comment}
\usepackage{appendix}
\usepackage{multirow}
\usepackage{url}
\usepackage{color}
\usepackage[normalem]{ulem}
\usepackage{cancel}

\newcommand{\jon}[2]{{#2}}
\newcommand{\joned}[1]{}

\newcommand{\incgraph}[3]{\includegraphics[angle=#1, width=#2\textwidth]{#3}}

\begin{document}

\title{Weighing The Evidence For A Gravitational-Wave Background\\In The First International Pulsar Timing Array Data Challenge}

\author{Stephen R. Taylor}
\email[email: ]{staylor@ast.cam.ac.uk}
\author{Jonathan R. Gair}
\email[email: ]{jgair@ast.cam.ac.uk}
\affiliation{Institute of Astronomy, \jon{}{University of Cambridge}, Madingley Road, Cambridge, CB3 0HA, UK }

\author{L. Lentati}
\email[email: ]{ltl21@cam.ac.uk}
\affiliation{Astrophysics Group, Cavendish Laboratory, JJ Thomson Avenue, Cambridge, CB3 0HE, UK}


\date{\today}

\begin{abstract}
We describe an analysis of the First International Pulsar Timing Array Data Challenge, which was designed to test the ability of new and existing algorithms to constrain the properties of a stochastic gravitational-wave background influencing the arrival-times of pulsar signals. We employ a robust, unbiased Bayesian framework developed by van Haasteren to study the three \textsc{Open} and \textsc{Closed} datasets of the IPTA data-challenge. We test various models for each dataset and use \textsc{MultiNest} to recover the evidence for the purposes of Bayesian model-selection. The parameter constraints of the favoured model are confirmed using an adaptive MCMC technique. Our results for \textsc{Closed}1 favoured a gravitational-wave background with strain amplitude at $f=1\text{ yr}^{-1}$, $A$, of $(1.1\pm 0.1)\times10^{-14}$, power spectral-index $\gamma=4.30\pm0.15$ and no evidence for red-timing noise or single-sources. The evidence for \textsc{Closed}2 favours a gravitational-wave background with $A=(6.1\pm 0.3)\times10^{-14}$, $\gamma=4.34\pm 0.09$ with no red-timing noise or single-sources. Finally, the evidence for \textsc{Closed}3 favours the presence of red-timing noise and a gravitational-wave background, with no single-sources. The properties of the background were $A=(5\pm 1)\times10^{-15}$ and $\gamma=4.23\pm0.35$, while the properties of the red-noise were $N_{\rm red}=(12\pm 4)$ ns and $\gamma_{\rm red}=1.5\pm0.3$. In all cases the redness of the recovered background is consistent with a source-population of inspiraling supermassive black-hole binaries. We also investigate the effect that down-sampling of the datasets has on parameter constraints and run-time. Finally we provide a proof-of-principle study of the ability of the Bayesian framework used in this paper to reconstruct the angular correlation of gravitational-wave background induced timing-residuals, comparing this to the Hellings and Downs curve.
\end{abstract}


\maketitle

\section{Introduction}
There is a large international effort focussed towards the first direct detection of gravitational waves (GWs). The existing and planned ground-based instruments, such as  AdLIGO \citep{AdvLIGO}, AdVirgo \citep{AdvVirgo} and KAGRA \citep{kagra2012}, are kilometre-scale interferometers sensitive to frequencies $\sim10-10^3$ Hz, and will likely be operating at design sensitivity by the end of this decade. In addition, it is hoped that a space-based interferometer with arm-lengths of $\sim10^9$ m, such as eLISA/NGO \citep{elisa-ngo}, will be operable by the end of 2020s, and sensitive in the range $\sim 0.1-100$ mHz.

The first indirect confirmation of the existence of GWs came from precision timing of the pulsar PSR B1913+16 \citep{taylor-weisberg-1989}, whose inferred binary orbital-energy loss was found to be consistent with the prediction of general relativity. This precision analysis was made possible by the often sub-$\mu$s level of timing precision achieved through the measurements of pulsar-signal time-of-arrivals (TOAs) \citep{pulsar-precision-1,pulsar-precision-2} whose accuracy can rival that of atomic clocks.

The precision of millisecond pulsars can be exploited for GW detection through the use of pulsar timing arrays (PTAs) \citep{foster-backer-1990}, which can effectively use the Milky Way as a kpc-scale GW-detector. Tens of Galactic millisecond pulsars have been observed over several years to search for the influence of a GW perturbing the space-time metric along each pulsar-Earth line-of-sight \citep{sazhin-1978,detweiler-1979,estabrook-1975,burke-1975}. PTAs are sensitive to low GW frequencies ($1-10$ nHz), where this range is set by the observational time-span ($f_{\rm low}\sim1/T$) and the cadence ($f_{\rm high}\sim1/(2\Delta T)$), and as such are complementary to other GW detection experiments.

It is not only GWs which can induce deviations of the TOAs from a timing model. The dominant perturbation is caused by the deterministic spin-down of the pulsar itself, as its rotational energy is extracted to power the EM outflow. There are also stochastic contributions to the deviations caused by a variety of sources, including clock noise, receiver noise and variations of the dispersion measure of the intervening interstellar medium. These effects must be accounted for and removed from the TOAs to produce the timing residuals, which then contain only the influence of unmodelled phenomena, including GWs. While there is a rich literature on the subject of the detection of single GW-sources using pulsar-timing (e.g.,\ \citep{sesana-vecchio-volonteri-2009,babak-sesana-2012,sesana-vecchio-2010,lee-wex-2011,sesana-vecchio-colacino-2008,petiteau-ga-2012}), the most likely source of GWs for PTAs is a stochastic gravitational-wave background (GWB).

An isotropic, stochastic GWB may be created by a superposition of many single sources which are not individually resolvable. In the PTA band, the largest contribution will likely come from a cosmological population of inspiraling supermassive-black-hole-binary (SMBHB) systems, with typical masses $\sim 10^4 - 10^{10}$ M$_{\odot}$. 

The fractional energy-density of the Universe in a GW-background is usually given as,
\begin{equation}
\Omega_{\rm GW}(f) = \frac{1}{\rho_{\rm{c}}}\frac{d\rho_{\rm GW}(f)}{d(\ln f)} = \frac{\pi}{4}f^2h_c(f)^2,
\end{equation}
where $f$ is the observed GW-frequency, $\rho_{\rm{c}}=3H^2/8\pi G$ is the energy-density required for a flat Universe, and $h_c(f)$ is the characteristic strain of the GW-background in a frequency interval centred at $f$.

The characteristic strain spectrum of a GW-background resulting from inspiraling binary systems is approximately $h_c(f)\propto f^{-2/3}$ \citep{begelman1980,phinney2001,jaffe-backer-2003,wyithe-loeb-2003}. We can approximate the characteristic strain spectrum of a GW-background from other sources as a power-law also. Some measurable primordial background contributions may have a power-law index of $-1$ \citep{grishchuk-1976,grishchuk-2005}, while the background from decaying cosmic strings \citep{vilenkin-1981a,vilenkin-1981b,olmez-2010,sanidas-2012} may have $-7/6$ \citep{damour-vilenkin-2005}. For most models of interest, we can describe an isotropic, stochastic GW-background by \citep{jenet-2006},
\begin{equation}
h_c(f) = A\left(\frac{f}{\rm{yr}^{-1}}\right)^{\alpha}.
\end{equation}
This characteristic strain spectrum is related to the one-sided power spectral density of the induced timing residuals by,
\begin{equation}
S(f) = \frac{1}{12\pi^2}\frac{1}{f^3}h_c(f)^2 = \frac{A^2}{12\pi^2}\left(\frac{f}{\rm{yr}^{-1}}\right)^{-\gamma}\rm{yr}^3,
\end{equation} 
where $\gamma\equiv3-2\alpha$.

\citet{hellings-downs-1983} developed a simple cross-correlation technique for pulsars affected by the same stochastic, isotropic GWB, showing that the cross-correlation of the induced timing-residuals has a distinctive angular signature dependant only on the angular-separation of the pulsars:
\begin{equation} \label{eq:hell-down}
\zeta_{ab} = \frac{3}{2}x\ln(x) - \frac{1}{4}x + \frac{1}{2} + \frac{1}{2}\delta_{ab},
\end{equation}
where $x = (1-\cos\theta_{ab})/2$, and $\theta_{ab}$ is the angular separation of the pulsar sky-locations.

While a GWB induces correlated residuals which typically have a steep, red spectrum, there may be additional uncorrelated red-noise contributions from rotational irregularities in each individual pulsar \citep{shannon-cordes-2010}, for which the spectrum is given as,
\begin{equation}
S(f) = N_{\rm red}^2\left(\frac{1}{1\text{yr}^{-1}}\right)\left(\frac{f}{1\text{yr}^{-1}}\right)^{-\gamma_{\rm red}}.
\end{equation}
Over the last several years constraints on the amplitude of an isotropic GWB have been published by the three major PTA collaborations \citep{demorest-2012,yardley-2011,van-haasteren-limits-2011}, the European Pulsar Timing Array (EPTA) \citep{epta-site}, the North American Nanohertz Observatory for Gravitational Waves (NANOGrav) \citep{nanograv-site}, and the Parkes Pulsar Timing Array (PPTA) \citep{ppta-site}. The International Pulsar Timing Array (IPTA) \citep{ipta-site} consortium combines these three efforts and recently initiated the first ``IPTA Data Challenge'' \citep{first-ipta-challenge} whose aim was to test new and existing algorithms for the purpose of constraining the properties of a background of GWs using PTAs.\footnote{Full details of this first challenge will be made available in a forthcoming IPTA data-challenge paper.}

In this paper we describe an analysis of the first IPTA data challenge. We use the time-domain Bayesian framework developed by \citet{van-haasteren-limits-2011},\citep{van-haasteren-levin-2012}. Although faster implementations of this method have recently been suggested \citep{van-haasteren-compression-2012,lentati-spectrum-2012}, we employ the original framework and analyse the uncompressed data for the purposes of model-selection with the recovered Bayesian evidence. To evaluate the evidence of posterior parameter distributions, we use the \textsc{MultiNest} nested sampling algorithm, and confirm favoured-model parameter constraints with an adaptive MCMC algorithm.

This paper is organised as follows. In Sec.\ \ref{sec:bayesian-inference} we describe Bayesian inference and how the evidence, which is the key quantity for model-selection, is computed. Section \ref{sec:pulsar-timing-analysis} describes pulsar timing analysis, including how the raw data is processed and a derivation of the likelihood of the timing-residuals given the timing-model and GWB parameters. The stochastic sampling techniques we have employed are described in Sec.\ \ref{sec:stochastic-sampling-techniques}, followed by a description of the IPTA challenge data in Sec.\ \ref{sec:data-description}. Results are presented in Sec.\ \ref{sec:results}, followed in Sec.\ \ref{sec:accel-down-sample} by brief studies of acceleration of the algorithm via down-sampling of the data and using the data to recover the Hellings and Downs correlation curve. We finish with our conclusions in Sec.\ \ref{sec:conclusions}.

\section{Bayesian inference}\label{sec:bayesian-inference}
Bayes' theorem states that the \textit{posterior} probability density function (PDF), $p(\vec{{\mu}}|D,\mathcal{H})$, of the parameters $\vec{{\mu}}$ describing a hypothesis model $\mathcal{H}$, and given data $D$ is
\begin{equation} \label{eq:bayes-theorem}
p(\vec{{\mu}}|D,\mathcal{H}) = \frac{p(D|\vec{{\mu}},\mathcal{H})p(\vec{{\mu}}|\mathcal{H})}{p(D|\mathcal{H})},
\end{equation}
where,
\begin{align}
p(D|\vec{{\mu}},\mathcal{H})&\equiv\mathcal{L}(\vec\mu)=\text{likelihood of data given parameters,}\nonumber\\
p(\vec{{\mu}}|\mathcal{H})&\equiv\pi(\vec\mu)=\text{prior PDF of parameters,}\nonumber\\
p(D|\mathcal{H})&=\mathcal{Z}=\text{Bayesian evidence.}
\end{align}

The Bayesian evidence, $\mathcal{Z}$, is the probability of the observed data given the model ${\cal H}$
\begin{equation}\label{eq:evidence}
\mathcal{Z} = \int \mathcal{L}(\vec\mu)\pi(\vec\mu)d^N\mu.
\end{equation}
For posterior inference within a model, ${\cal Z}$ plays the role of a normalisation constant and can be ignored. However, if we want to perform model selection then this evidence value becomes key. In Bayesian model comparison we compute the Bayes factor
\begin{equation}
\frac{p(\mathcal{H}_2|\vec D)}{p(\mathcal{H}_1|\vec D)} = \frac{p(\vec D|\mathcal{H}_2)p(\mathcal{H}_2)}{p(\vec D|\mathcal{H}_1)p(\mathcal{H}_1)}=\frac{\mathcal{Z}_2\times p(\mathcal{H}_2)}{\mathcal{Z}_1\times p(\mathcal{H}_1)}.
\end{equation}
where $p(\mathcal{H}_2)/p(\mathcal{H}_1)$ is the prior probability ratio for the two competing models. This can often be set to one, and we will do so in the remainder of this analysis. The Bayes factor is then just the evidence ratio. Since the evidence is the average of the likelihood over the prior volume, it automatically incorporates Occam's razor, which states that, all else being equal, a model with fewer parameters is favoured. Hypothesis ${\cal H}_1$ is chosen if the Bayes factor is sufficiently large. Jeffreys \citep{jeffreys1983} gave a scale interpretation for the Bayes factor, which is shown in Table \ref{tab:jeffreys-scale-bayes}.

We employ Bayesian model-selection in the following study to determine which phenomena provide the best explanation for the observed pulsar TOAs in the first IPTA data challenge.

\begin{table}
\caption{\label{tab:jeffreys-scale-bayes}An interpretation of the Bayes factor in determining which model is favoured, as given by Jeffreys \citep{jeffreys1983}.}
\begin{ruledtabular}
\begin{tabular}{c c l}
Bayes factor, $\mathcal{K}$ & $\ln(\mathcal{K})$ & Strength of evidence\\
\hline
$< 1:1$ & $< 0$ & Negative (supports $\mathcal{H}_1$)\\
$1:1$ to $3:1$ & $0 -1.1$ & Barely worth mentioning\\
$3:1$ to $10:1$ & $1.1 - 2.3$ & Substantial\\
$10:1$ to $30:1$ & $2.3 - 3.4$ & Strong\\
$30:1$ to $100:1$ & $3.4 - 4.6$ & Very strong\\
$> 100:1$ & $> 4.6$ & Decisive\\
\end{tabular}
\end{ruledtabular}
\end{table}

\section{Pulsar Timing Analysis}\label{sec:pulsar-timing-analysis}
Observations of pulsars lead to measurements of the pulsar TOAs. The emission-time of a pulse is given in terms of the observed TOA by \citep{tempo2-1,tempo2-2},
\begin{equation}
t_{\rm em}^{\rm psr} = t_{\rm arr}^{\rm obs} - \Delta_{\odot} - \Delta_{\rm IS} - \Delta_{\rm B},
\end{equation}
where $\Delta_{\odot}$ is the transformation from the site TOAs to the Solar-system barycentre, $\Delta_{\rm IS}$ accounts for the delaying-effects as the pulse propagates through the interstellar medium, and $\Delta_{\rm B}$ converts to the pulsar-frame for binary pulsars.

In the first IPTA data challenge \citep{first-ipta-challenge} the raw data is in the form of pulsar parameter files (``.par'') and timing files (``.tim''). The parameter file contains first estimates of the pulsar timing-model parameters; these parameters describe deterministic contributions to the arrival times. The vector of measured arrival times will be composed of a deterministic and a stochastic contribution (from time-correlated stochastic signals which are modelled by a random Gaussian process),
\begin{equation}
\vec{t}^{\rm{arr}} = \vec{t}^{\rm{det}} + \delta \vec{t}^{\rm{rgp}}.
\end{equation}
The stochastic process has auto-correlation,
\begin{equation}\label{eq:pre-fit-stoch-cov}
C_{ij}=\langle\delta t_i^{\rm{rgp}}\delta t_j^{\rm{rgp}}\rangle, 
\end{equation}
where the elements of the covariance matrix are parametrised by a set of parameters, $\vec\phi$. Using the Wiener-Khinchin theorem, we can then define the auto-correlation as the Fourier transform of the power spectral density,
\begin{equation}
C(\tau_{ij}) = \int_0^{\infty}S(f)\cos(f\tau_{ij})df,
\end{equation}
where $\tau_{ij}=2\pi\vert t_i - t_j \vert$, and $S(f)$ is the power spectral density of the time-series $\delta \vec{t}^{\rm{rgp}}$. A closed-form expression for the auto-correlation of a time-series influenced by an underlying power-law PSD is given in \citet{van-haasteren-limits-2011}, and is used in the following.

\subsection{Processing raw arrival-times}
The ``.par'' and ``.tim'' files are fed to the \textsc{Tempo2} software package \citep{tempo2-1,tempo2-2,tempo2-3} which processes the raw arrival-times. A vector of ``pre-fit'' timing-residuals are computed using the first guesses, $\beta_{0,i}$, of the ``$m$'' timing-model parameters from the ``.par'' files. This first guess is usually precise enough so that a linear approximation can be used in the TOA fitting procedure, so that the post-fit timing residual are
\begin{equation}
\delta \vec{t} = \delta \vec{t}^{\rm{prf}} + M\vec\xi,
\label{eq:postfit}
\end{equation}
where $\delta \vec{t}^{\rm{prf}}$ are the pre-fit timing-residuals (length $n$), $\vec\xi$ is the vector of deviations from the pre-fit parameters (length $m$) defined as $\xi_a = \beta_a - \beta_{0,a}$, and $M$ is the $(n\times m)$ ``design-matrix'', describing how the residuals depend on the timing-model parameters. \textsc{Tempo2} does not take into account the possible time-correlated stochastic signal in the TOAs, but performs a weighted least-squares fit for the timing-model parameter values. Hence it is possible that some of the time-correlated stochastic signal is absorbed in this fitting procedure, which is undesirable.

The \textsc{Tempo2} analysis provides output-residuals and the design matrix, $M$. The design matrix describes the dependence of the timing residuals on the timing-model parameters. The output-residuals form the input data vector for further study. 

\subsection{Generalised least-squares (GLS) estimator of stochastic and deterministic parameters}
We now want to use the \textsc{Tempo2} output-residuals to determine any correlated stochastic signal affecting the pulse arrival times. We assume that the part of the stochastic signal removed by the fitting procedure is small, so that the \textsc{Tempo2} output-residuals are related linearly to the stochastic contribution to the residuals 
\begin{equation}
\delta \vec{t} = \delta \vec{t}^{\rm{rgp}} + M\vec\xi,
\end{equation}
where, in this case, $\delta \vec{t}$ refers to the output-residuals from \textsc{Tempo2}. We note that the $\vec\xi$ appearing in this equation is different from that appearing in Eq.~(\ref{eq:postfit}).

The stochastic timing residuals, $\delta \vec{t}^{\rm rgp}$, arise from a time-correlated stochastic process with covariance matrix $C$. This covariance matrix may contain contributions from the GWB, white-noise from TOA-errors, and possibly red-timing noise which is uncorrelated between different pulsars. The likelihood of measuring post-fit residuals, $\delta \vec{t}$, given the fit parameters $\vec\xi$ and stochastic parameters, $\vec\phi$, is,
\begin{align}
\mathcal{L}(\delta\vec{t}\vert\vec{\xi},\vec{\phi}) =& \frac{1}{\sqrt{(2\pi)^n{\rm{det}} C}}\times\nonumber\\
&\exp{\left(-\frac{1}{2}\left(\delta\vec{t} - M\vec{\xi}\right)^{T}C^{-1}\left(\delta\vec{t} - M\vec{\xi}\right)\right)}.
\end{align}
This likelihood expression is effectively a GLS estimator, and is the basis for the framework used in this paper to study the first IPTA data challenge.

If we assume flat priors on the timing-model parameters then these parameters can be analytically marginalised over. The posterior distribution marginalised over timing-model parameters is \citep{van-haasteren-2011},
\begin{equation}\label{eq:long-vh-marge}
P(\vec\phi\vert\delta\vec{t})\propto\frac{1}{\sqrt{{\rm{det}}C\times{\rm{det}}(M^{T}C^{-1}M)}}\exp{\left(-\frac{1}{2}\delta\vec{t}^{T}C'\delta\vec{t}\right)},
\end{equation}
where $C' = C^{-1} - C^{-1}M\left(M^{T}C^{-1}M\right)^{-1}M^{T}C^{-1}$. When dealing with large datasets and many pulsars, $C'$ \jon{naturally}{} involves the multiplication and inversion of high dimensional matrices. In the case of multiple pulsars, the design matrix, covariance matrix and total residual vector are given by,
\begin{align}
C_{\rm{tot}} &= \begin{pmatrix}C_{11} & C_{12} & \ldots\\C_{21} & C_{22} & \ldots\\\vdots & \vdots & \ddots\end{pmatrix},\nonumber\\
M_{\rm{tot}} &= \begin{pmatrix}M_1 & & & \\ & M_2 & & \\ & & \ddots &\end{pmatrix}, \nonumber\\
\delta\vec{t}_{\rm{tot}} &=  \begin{pmatrix}\delta\vec{t}_1 \\ \delta\vec{t}_2 \\ \vdots\end{pmatrix},
\end{align}

where $C_{ab}$ is the auto-covariance matrix between pulsars $a$ and $b$, $M_a$ are the individual pulsar design matrices and $\delta\vec{t}_a$ are the individual pulsar residual vectors.  We can split $C_{ab}$ into contributions from various stochastic sources. So, the covariance between the $i$th residual of pulsar $a$ and the $j$th residual of pulsar $b$ is,
\begin{equation}
C_{(ai)(bj)} = C^{\rm GW}_{(ai)(bj)} + C^{\rm TOA}_{(ai)(bj)} + C^{\rm EQUAD}_{(ai)(bj)} + C^{\rm RN}_{(ai)(bj)},
\end{equation}
where,
\begin{widetext}
\begin{align}
C^{\rm GW}_{(ai)(bj)} =& \frac{A^{2}}{12\pi^{2}}\zeta_{ab}\left(\frac{1\rm{ yr}^{-1}}{f_{l}}\right)^{\gamma-1}\left[\Gamma(1-\gamma)\sin{\left(\frac{\pi\gamma}{2}\right)}(f_{l}\tau_{ij})^{\gamma-1}-\displaystyle\sum_{n=0}^{\infty}(-1)^n\frac{\left(f_{l}\tau_{ij}\right)^{2n}}{(2n)!(2n+1-\gamma)}\right],\nonumber\\
C^{\rm TOA}_{(ai)(bj)} =& (\text{EFAC}_a)^2\Delta t_{(ai)}^2\delta_{ab}\delta_{ij},\nonumber\\
C^{\rm EQUAD}_{(ai)(bj)} =& (\text{EQUAD}_a)^2\delta_{ab}\delta_{ij},\nonumber\\
C^{\rm RN}_{(ai)(bj)} =& N_{{\rm red},a}^2\delta_{ab}\left(\frac{1\rm{ yr}^{-1}}{f_{l}}\right)^{\gamma_{\rm red}-1}\left[\Gamma(1-\gamma_{\rm red})\sin{\left(\frac{\pi\gamma_{\rm red}}{2}\right)}(f_{l}\tau_{ij})^{\gamma_{\rm red}-1}-\displaystyle\sum_{n=0}^{\infty}(-1)^n\frac{\left(f_{l}\tau_{ij}\right)^{2n}}{(2n)!(2n+1-\gamma_{\rm red})}\right].
\end{align}
\end{widetext}
In the order listed, these are i) the GW-background covariance, $C^{\rm GW}_{(ai)(bj)}$,which depends on $\zeta_{ab}$, the Helling-Downs correlation between pulsars $a$ and $b$; ii) the TOA error-bar covariance, $C^{\rm TOA}_{(ai)(bj)}$, arising from white noise in each individual pulsar, which is characterised by a separate, pre-specified and fixed, amplitude $\Delta t_{(ai)}$ for each pulsar $a$ and time $i$, plus an overall scaling factor, EFAC, which is common to all pulsars but is allowed to vary as a model parameter; iii) the covariance, $C^{\rm EQUAD}_{(ai)(bj)}$, of an additional white-noise which is common to all pulsars and time-stamps and characterised by a single amplitude parameter EQUAD; and iv) the covariance, $C^{\rm RN}_{(ai)(bj)}$, of a red timing-noise in each pulsar, which is modelled as a power-law with amplitude $N_{\rm red}$ and slope $\gamma_{\rm red}$. The timing-model fit reduces the sensitivity of the residuals to the low-frequency cutoff, $f_l$, which is required when $\gamma\geq1$. Hence, provided $f_lT<<1$ we can ignore all terms with $n\geq2$ in the infinite summation. To avoid numerical artefacts we choose $f_l = 10^{-3}$ yr$^{-1}$.

Expression (\ref{eq:long-vh-marge}) can be written more compactly and in a way which is slightly faster to compute \citep{van-haasteren-levin-2012}:
\begin{align}\label{eq:vh-marg}
P(\vec\phi\vert\delta\vec{t})&=\pi(\vec\phi)\times\frac{1}{\sqrt{(2\pi)^{n-m}{\rm{det}}(G^{T}CG)}} \nonumber\\
&\quad\exp{\left(-\frac{1}{2}\delta\vec{t}^{T}G\left(G^{T}CG\right)^{-1}G^{T}\delta\vec{t}\right)},
\end{align}
where $G$ is the matrix constructed from the final $(n-m)$ columns of the matrix $U$ in the SVD of the design matrix, $M=U\Sigma V^*$. The matrix $G$ can be pre-computed and stored in memory for use in each likelihood calculation. 

Equation (\ref{eq:vh-marg}) provides a robust, unbiased Bayesian framework for the search for correlated signals in PTAs, and is used in all of the following analysis.

\subsection{Including single-sources in the search}

While this analysis focuses on the detection and characterisation of a background of GWs, we will also consider the possible presence of a single monochromatic or burst source perturbing the arrival times of pulses. In practice, it may be necessary to include single-source models in all background searches to allow the background-induced residuals to be described by Gaussian statistics \citep{ravi-2012}. There is a large literature on the subject of single-source detection in the context of PTAs. In particular, \citet{sesana-vecchio-colacino-2008} describe how the commonly used approximation of a single power-law spectrum for a stochastic background from inspiraling SMBHBs breaks down at frequencies higher than $10^{-8}$ Hz due to the dominance of single sources. Likewise, the authors of \citep{sesana-vecchio-volonteri-2009,babak-sesana-2012,sesana-vecchio-2010,lee-wex-2011,petiteau-ga-2012} have studied the ability of a PTA to infer the presence of multiple resolvable monochromatic sources, and to constrain their properties.

We use the formalism of \citep{van-haasteren-2011} to combine the search for single sources with the search for a background. The formalism is a simple modification to the background search, in which the residuals are now described by,
\begin{equation}
\delta\vec{t} = \delta\vec{t}^{\rm rgp} + \vec{s} + M\vec\xi,
\end{equation}
where $\vec{s}$ is the deterministic contribution to the residuals from a single source.

Models for the $+$,$\times$ GW polarisation amplitudes can be used to compute the frequency-shift of pulses induced by the GW. The redshift of signals from a pulsar in the direction of unit vector $\hat{p}$, induced by the passage of a GW coming from direction $\hat{\Omega}$ is,
\begin{equation}\label{eq:pulse-freq-shift}
z(t,\hat\Omega) = \frac{1}{2}\frac{\hat{p}^i\hat{p}^j}{1+\hat\Omega\cdot\hat{p}}\Delta h_{ij},
\end{equation}
where $\Delta h_{ij}\equiv h_{ij}(t_p,\hat\Omega)-h_{ij}(t_e,\hat\Omega)$, is the difference in the metric perturbation at the pulsar and at the solar system barycentre. We ignore the uncorrelated pulsar-term in this analysis.

This frequency-shift is \jon{then}{} integrated over time to give the induced timing residuals,
\begin{equation}\label{eq:induced-residuals}
R(t)\equiv\int_0^tz(t')dt'.
\end{equation}
We will search for monochromatic and burst sources and descriptions of the models used to describe these sources are given in Appendix \ref{sec:single-source-description}.

\section{Stochastic sampling techniques}\label{sec:stochastic-sampling-techniques}
We now discuss two different stochastic sampling techniques used to reconstruct the posterior PDF of the model parameters and to calculate the evidence value for Bayesian model-selection.

\subsection{Markov chain Monte Carlo (MCMC) sampling}
Markov chain Monte Carlo (MCMC) techniques provide an efficient way to explore a model parameter space. An initial point, $\vec{x_0}$, is drawn from the \textit{prior} distribution and then at each subsequent iteration, $i$, a new point, $\vec{y}$, is drawn from a \textit{proposal distribution}, $q({\vec{y}}|{\vec{x}})$ and the Metropolis-Hastings ratio evaluated,
\begin{equation}
R=\frac{{\pi}(\vec{y}){\mathcal{L}}(D|{\vec{y}},\mathcal{H})q({\vec{x_i}}|\vec{y})}{{\pi}(\vec{x_i}){\mathcal{L}}(D|{\vec{x_i}},\mathcal{H})q({\vec{y}}|{\vec{x_i}})}.
\end{equation}
A random sample is drawn from a uniform distribution, $u\in U[0,1]$, and if $u<R$ the move to the new point is accepted and we set ${\vec{x}_{i+1}}={\vec{y}}$. If $u>R$, the move is rejected and we set ${\vec{x}_{i+1}}={\vec{x_i}}$. 

The MCMC samples can be used to carry out integrals over the posterior
\begin{equation}
\int f(\vec{x})p(\vec{x}|D,\mathcal{H})d\vec{x}\approx{\frac{1}{N}}\displaystyle\sum_{i=1}^Nf(\vec{x_i}).
\end{equation}
The $1$D marginalised posterior probability distributions in individual model parameters then follow by binning the chain samples in that parameter.

The trick to using this technique efficiently is to choose an appropriate proposal distribution. In our analysis we employ an adaptive MCMC procedure, which utilises an `in-flight' estimation of the sampled-chain's covariance matrix to construct an updating proposal distribution. This covariance matrix is updated at each iteration, with a certain chain memory \citep{haario1999,haario2001,dunkley2005}. We use several of the procedures outlined in \citep{dunkley2005}. A full description of this technique can be found in \citet[][and references therein]{taylor-gair-et-2012}.

A single likelihood calculation on one core can take as much as $\sim40$ seconds, so it is very important that we achieve a fast burn-in. We do this for each dataset by using the built-in threading capability of \textsc{LAPACK} \citep{lapack-site}. This allows as many as $12$ cores to speed-up a single matrix multiplication operation, and ultimately reducing the likelihood calculation time to $\sim5$ seconds. Hence we use $5$ independent chains with threading to burn-in. After approximately $2$ hours burn-in is achieved, and we can sample from the end of these chains to initiate a larger run of $512$ cores, with no threading, to collect posterior samples. Collecting $\sim10^5$ samples takes approximately $3$ hours.

\subsection{Nested Sampling \& \textsc{MultiNest}}
The nested sampling algorithm is a Monte Carlo method, originally proposed by Skilling \citep{skilling2004} for evaluating the Bayesian evidence, $\mathcal{Z}$. For a full description of the \textsc{MultiNest} algorithm see \citet{feroz2008,feroz2009}, but we describe the basics in the following section.     

The basic idea is to populate parameter space with ``live'' points drawn from the prior. These points move as the algorithm proceeds, climbing together through nested contours of increasing likelihood. At each iteration, the points are ordered in terms of their likelihood, and the point with lowest likelihood is replaced by a point with higher likelihood than this lowest-likelihood point.

The biggest difficulty in nested sampling is to efficiently sample points of higher likelihood to allow the live-points to climb. If we were to simply draw points from the prior volume, then the acceptance rate of new points in the live-set would steadily decrease, since at later iterations the live-set occupies a smaller and smaller volume of the prior space as it climbs. \textsc{MultiNest} overcomes this drawback by using a sophisticated ellipsoidal rejection-sampling technique, whereby the current live-set is enclosed by (possibly overlapping) ellipsoids, and a new point drawn uniformly from the enclosed region. This technique successfully copes with multimodal distributions and parameter spaces with strong, curving degeneracies.

The evidence is calculated by transforming the multi-dimensional integral in Eq.\ (\ref{eq:evidence}) into a one-dimensional integral which is easily numerically evaluated. We define the prior volume, $X$ as,
\begin{equation}
dX = \pi(\vec\mu)d^N\mu,
\end{equation}
such that,
\begin{equation}
X(\lambda) = \int_{\mathcal{L}(\vec\mu)>\lambda}\pi(\vec\mu)d^N\mu,
\end{equation}
where the integral extends over the region of the $N$-dimensional parameter space contained within the iso-likelihood contour $\mathcal{L}(\vec\mu)=\lambda$. Hence, Eq.\ (\ref{eq:evidence}) can be written as,
\begin{equation}
\mathcal{Z} = \int_0^1\mathcal{L}dX,
\end{equation}
where $\mathcal{L}(X)$ is a monotonically decreasing function of $X$. If we order the $X$ values ($0<X_M<\ldots<X_1<X_0=1$), then the evidence, $\mathcal{Z}$ can be approximated numerically using the simple trapezium rule,
\begin{equation}
\mathcal{Z} = \displaystyle\sum_{i=1}^M\mathcal{L}_iw_i,
\end{equation}
where the weights, $w_i$, are given by $w_i = \left(X_{i-1}-X_{i+1}\right)/2$.

As a by-product of the exploration of the parameter space by the evolving live-set, \textsc{MultiNest} also permits reconstruction of the parameter posterior PDFs. Once $\mathcal{Z}$ is found, the final live-set, as well as the discarded points, are collected and assigned probability weights to give the posterior probability of each point. These points can be binned to give full and marginalised posterior PDFs.

With \textsc{MultiNest}'s built-in MPI routines we use $\sim800$ live-points for all runs, and typically employ $160$ cores such that \textsc{MultiNest} finishes in less than $12$ hours.

\section{Description of data}\label{sec:data-description}

The first IPTA data challenge consists of three ``\textsc{Open}'' and three ``\textsc{Closed}'' datasets. For the ``\textsc{Open}'' datasets, the properties of the noise and background were given, allowing the calibration and testing of algorithms. The parameters of the ``\textsc{Closed}'' datasets are not due to be revealed until after the deadline. However the format of the data is the same in both sections. In all data sets, we have $36$ pulsars distributed across the sky, each having $130$ pulse arrival times measured over an average time span of $5$ years.

\subsection{\textsc{Open} data}
The open section of the challenge consisted of three separate sets of data, increasing in the level of complexity. All three data sets contained a GWB background with a power spectral density slope given by $\gamma\equiv 3-2\alpha=13/3$, which is consistent with a background induced by a population of inspiraling SMBHBs. Each data set was for a total time span of $5$ years with $130$ observations per pulsar. The data sets differed in their sampling cadence and in the TOA noise in the pulsars.

In ``\textsc{Open}1'' the intrinsic noise in the pulsar TOAs was white, with the same amplitude of $100$ ns in each pulsar, and the sampling cadence was uniform with a rate of one sample every two weeks. The characteristic strain-spectrum amplitude of the GWB at $f=1\text{ yr}^{-1}$ was $A=5\times 10^{-14}$.

In ``\textsc{Open}2'' the intrinsic TOA noise was again white, but the amplitude was different for each pulsar,with nominal values as given in Table \ref{tab:open2-white-noise}. These white noise levels are consistent with realistic levels measured for IPTA pulsars. In addition, the sampling rate was no longer uniform but random, with an average cadence of $2$ weeks $\pm 5$ days.The characteristic strain-spectrum amplitude of the GWB was $A=5\times 10^{-14}$.

\begin{table*}
\caption{\label{tab:open2-white-noise}The white noise levels of each pulsar in the \textsc{Open}2 and \textsc{Open}3 datasets \citep{ipta-wn-levels}.}
\begin{ruledtabular}
\begin{tabular}{c c c c c c c c}
Pulsar & RMS WN ($\mu$s) & Pulsar & RMS WN ($\mu$s) & Pulsar & RMS WN ($\mu$s) & Pulsar & RMS WN ($\mu$s)\\
\hline
J0030+0451    &      0.31 & J1022+1001       &    0.37 & J1730-2304    &       0.83 & J1910+1256     &      0.17\\
J0218+4232    &       4.81 & J1024-0719        &    0.25 & J1732-5049     &       1.74 & J1918-0642    &       0.87\\
J0437-4715    &         0.03 & J1045-4509    &        2.68 & J1738+0333   &       0.24 & J1939+2134   &       0.02\\
J0613-0200     &      0.45 & J1455-3330    &        1.60 & J1741+1351    &        0.19 & J1955+2908   &      0.18\\
J0621+1002      &     9.58 & J1600-3053     &      0.23 & J1744-1134     &       0.14 & J2019+2425   &      0.66\\
J0711-6830       &     1.32 & J1603-7202    &       0.70 & J1751-2857    &        0.90 & J2124-3358   &       1.52\\
J0751+1807        &   0.78 & J1640+2224     &     0.19 & J1853+1303   &       0.17 & J2129-5721   &       0.87\\
J0900-3144     &      1.55 & J1643-1224      &      0.53 & J1857+0943   &       0.25 & J2145-0750    &     0.40\\
J1012+5307      &     0.32 & J1713+0747       &    0.04 & J1909-3744    &       0.04 & J2317+1439   &     0.25\\ 
\end{tabular}
\end{ruledtabular}
\end{table*}

In ``\textsc{Open}3'' the intrinsic pulsar noise has both a white component with levels as in ``\textsc{Open}2'' and an uncorrelated red component, which had the same power spectrum for each pulsar, but with a different realisation for each pulsar's TOAs. The red-noise power spectrum is $S(f)=5.77\times 10^{-22}\text{sec}^{1.3}f^{-1.7}$, where $f$ is in Hz. This is equivalent to the expression,
\begin{equation}
S(f) = N_{\rm red}^2\left(\frac{1}{1\text{yr}^{-1}}\right)\left(\frac{f}{1\text{yr}^{-1}}\right)^{-\gamma_{\rm red}},
\end{equation}
where $\gamma_{\rm red}=1.7$ and $N_{\rm red}=10.1$ ns. The characteristic strain-spectrum amplitude of the GWB was $A=10^{-14}$.

\section{Results}\label{sec:results}
We used \textsc{MultiNest} to compute evidence values for model selection and confirmed parameter constraints for the favoured model with adaptive MCMC. Except in certain circumstances which we will discuss, we made no attempt to recover the individual white-noise levels of each pulsar, nor have we permitted the pulsars in \textsc{Open}/\textsc{Closed}3 to have individual red-noise amplitude/index parameters. Rather, where necessary, pulsars share a global TOA error-bar scaling (GEFAC), extra white-noise contribution (GEQUAD), red-noise amplitude and red-noise index. However, all white-noise and red-noise is uncorrelated between different pulsars.

From experience with the open datasets, we found that uniform priors on all GWB and red-noise parameters was reasonable. A test-run on \textsc{Open}1 with uniform priors on the log of the GWB-amplitude yielded similar results. The posterior distribution was narrow compared to the prior which covered eight decades in amplitude, which explains why the results of uniforms priors on amplitudes and log-amplitudes were similar. When fitting for the EQUAD or white-noise parameters, our prior was $\log_{10}\left(\sigma_{\rm EQUAD/WN}\right)\in\left[-11.0,-3.0\right]$. To allow for the influence of prior choice on the evidence calculation, we only consider a model disfavoured if the evidence is lower than another model by at least $\Delta\ln(\mathcal{Z})=3$, which was the typical difference seen when different prior choices were tested. However, if two models have similar evidence but one has many more unconstrained parameters, we naturally favour the simpler model. 

Tables \ref{tab:all-open-models} and \ref{tab:all-closed-models} show all tested models for all datasets with the associated evidence. The strength with which certain models are favoured can be gauged using Table \ref{tab:jeffreys-scale-bayes}. We quote maximum-a-posteriori values and $1\sigma$ errors (half-width of $68\%$ credible region) for each parameter. All plots were made with GetDist in the CosmoMC package \citep{cosmomc-site}.

\subsection{\textsc{Open} Datasets}

\subsubsection{\textsc{Open}1}
\begin{figure}
  \incgraph{0}{0.5}{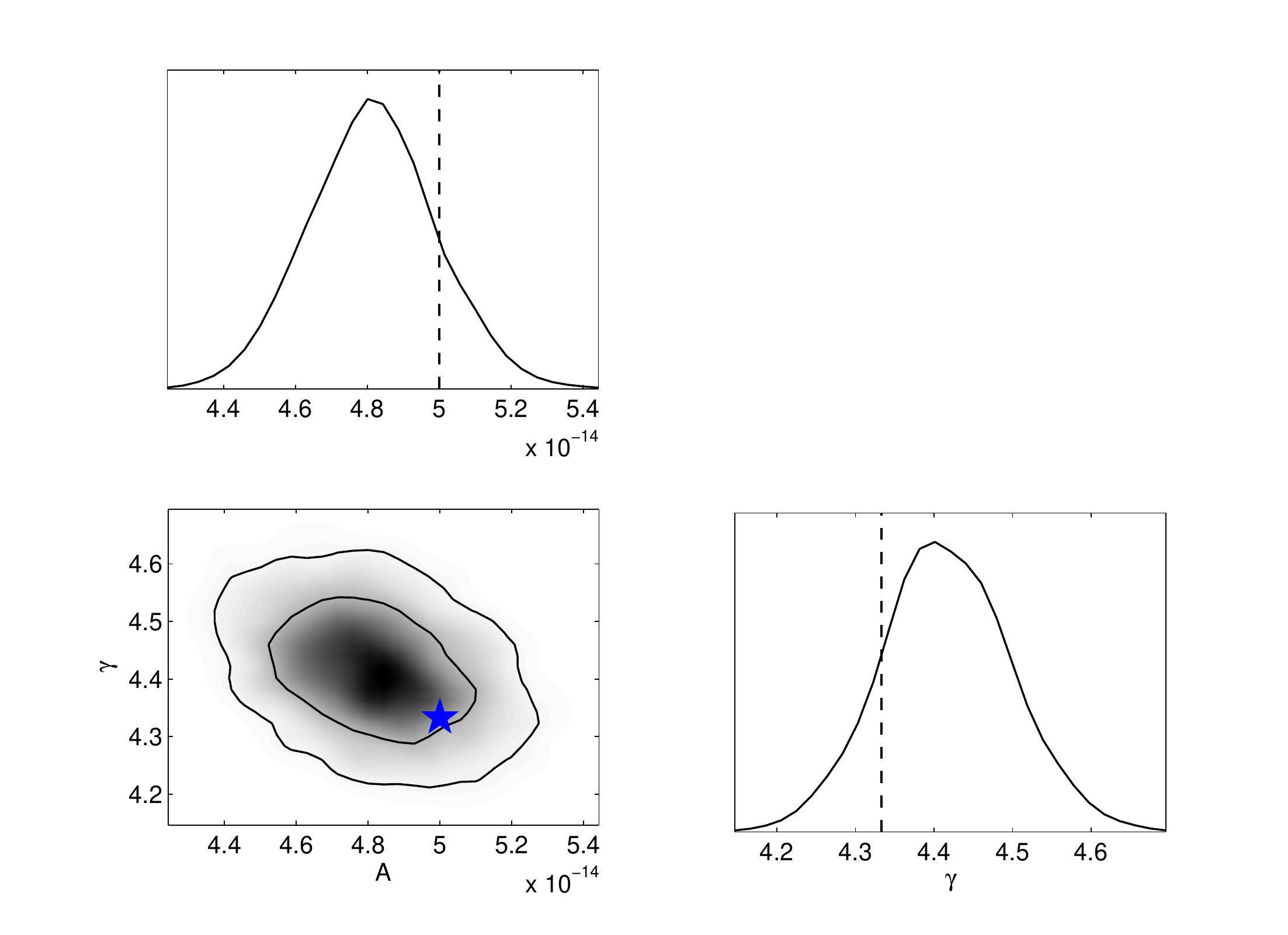}
  \caption{\label{fig:open1}The $1$D and $2$D marginalised posterior distributions of \textsc{Open}1. The star and dashed lines show the injected values of parameters.}
\end{figure}

This dataset provided the first test of the analysis pipeline. The time-stamp of each data sample and the white noise level in the pulsar were read in as input, along with the post-fit residuals. Although these were constant for all pulsars in \textsc{Open}1, this implementation allowed easy generalisation to the other data sets. The timing-models of all pulsars converged after a single \textsc{Tempo}2 fitting procedure.

From Table \ref{tab:all-open-models} it is clear that the evidence decisively favours the presence of a GWB. The properties of this background are $A=(4.8\pm 0.2)\times 10^{-14}$ and $\gamma=4.42\pm 0.08$. Figure \ref{fig:open1} shows the $1$D and $2$D marginalised posterior distributions for these parameters. These recovered parameters are consistent with the injected values. 

\begin{table}\scriptsize
\caption{\label{tab:all-open-models}Models tested for all \textsc{Open} datasets. The most favoured model for each dataset is shown in bold. The acronyms correspond to: Gravitational-Wave Background (GWB), Time-Of-Arrival (TOA), White Noise (WN), Global EFAC (GEFAC), Global EQUAD (GEQUAD) and Red-Noise (RN).}
\begin{ruledtabular}
\begin{tabular}{l c}
Model & $\ln(\mathcal{Z})$ \\
\hline
\multicolumn{2}{c}{\textsc{Open}1}\\
\hline
Null (TOA-errors only) & $-190502$\\
{\bf GWB, TOA-errors} & $\mathbf{61405.2\pm 0.1}$\\
\hline
\multicolumn{2}{c}{\textsc{Open}2}\\
\hline
Null (TOA-errors only) & $-821338$\\
GWB, TOA-errors & $55466.2\pm 0.1$\\
GWB, GEFAC, GEQUAD, TOA-errors & $55471.7\pm 0.1$\\
{\bf GWB, WN(J1857+0943)} & $\mathbf{55574.8\pm 0.1}$\\
GWB, WN(J0437-4715, J1857+0943, & $55568.2\pm 0.2$\\
\hspace{1.5cm}J1909-3744) &\\
WN(J1857+0943) & $-819198$\\
\hline
\multicolumn{2}{c}{\textsc{Open}3}\\
\hline
Null (TOA-errors only) & $37379$\\
GWB, TOA-errors & $56038.9\pm 0.1$\\
{\bf GWB, RN, TOA-errors} & $\mathbf{56037.8\pm 0.1}$\\
\end{tabular}
\end{ruledtabular}
\end{table}

\subsubsection{\textsc{Open}2}
\begin{figure*}
  \incgraph{0}{1.0}{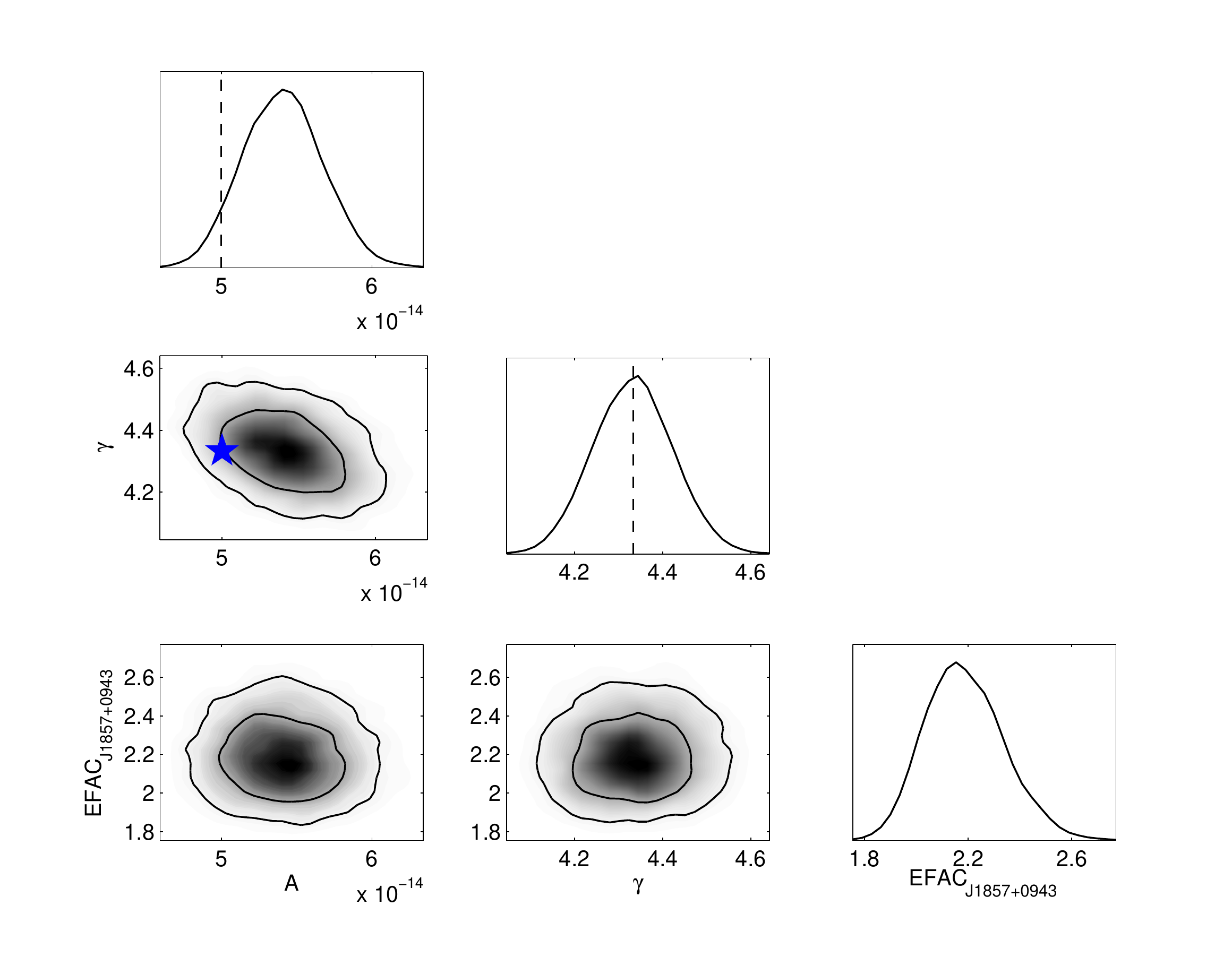}
  \caption{\label{fig:open2}The $1$D and $2$D marginalised posterior distributions of \textsc{Open}2. The star and dashed lines show the injected values of parameters.}
\end{figure*}

The differences between this dataset and \textsc{Open}1 were the random observation-cadence, and different white intrinsic TOA errors in each pulsar. These provided no difficulties for the pipeline.

From Table \ref{tab:all-open-models} it is clear that the evidence decisively favours the presence of a GWB, although the recovered background-amplitude showed a $2\sigma$ deviation from the injected value. Adding in GEFAC and GEQUAD parameters improved the evidence and reported $\text{GEFAC}=1.06\pm 0.01$ and GEQUAD consistent with zero. This was revisited after the analysis of \textsc{Closed}2 described below. Visual inspection of the output-residuals showed ``glitchy'' behaviour in J1857+0943, and reprocessing the arrival-times by \textsc{Tempo}2 indicated that the timing-solution had not converged after one fitting-iteration. Rather than re-fit we allowed the total white-noise in this pulsar to be a model parameter to be fitted. This created a huge improvement in the evidence and the recovery of the injected GWB parameters.

Fitting for the white-noise of several other pulsars which showed ``glitchy'' behaviour shifted the max-a-posteriori background-amplitude slightly closer to the injected value, but the evidence was slightly worse. We thus conclude that J1857+0943 was the main source of the inconsistency between the injected and recovered background-amplitude. This was the only pulsar whose timing-solution did not converge after a single \textsc{Tempo}2 fitting procedure. The recovered properties of the GWB are $A=(5.4\pm 0.3)\times10^{-14}$ and $\gamma=4.33\pm0.09$, with an effective EFAC on J1857+0943 of $2.2\pm 0.2$. Figure \ref{fig:open2} shows the $1$D and $2$D marginalised posterior distributions for these parameters. The recovered GWB properties are consistent with the injected values.

\subsubsection{\textsc{Open}3}
\begin{figure*}
  \incgraph{0}{1.0}{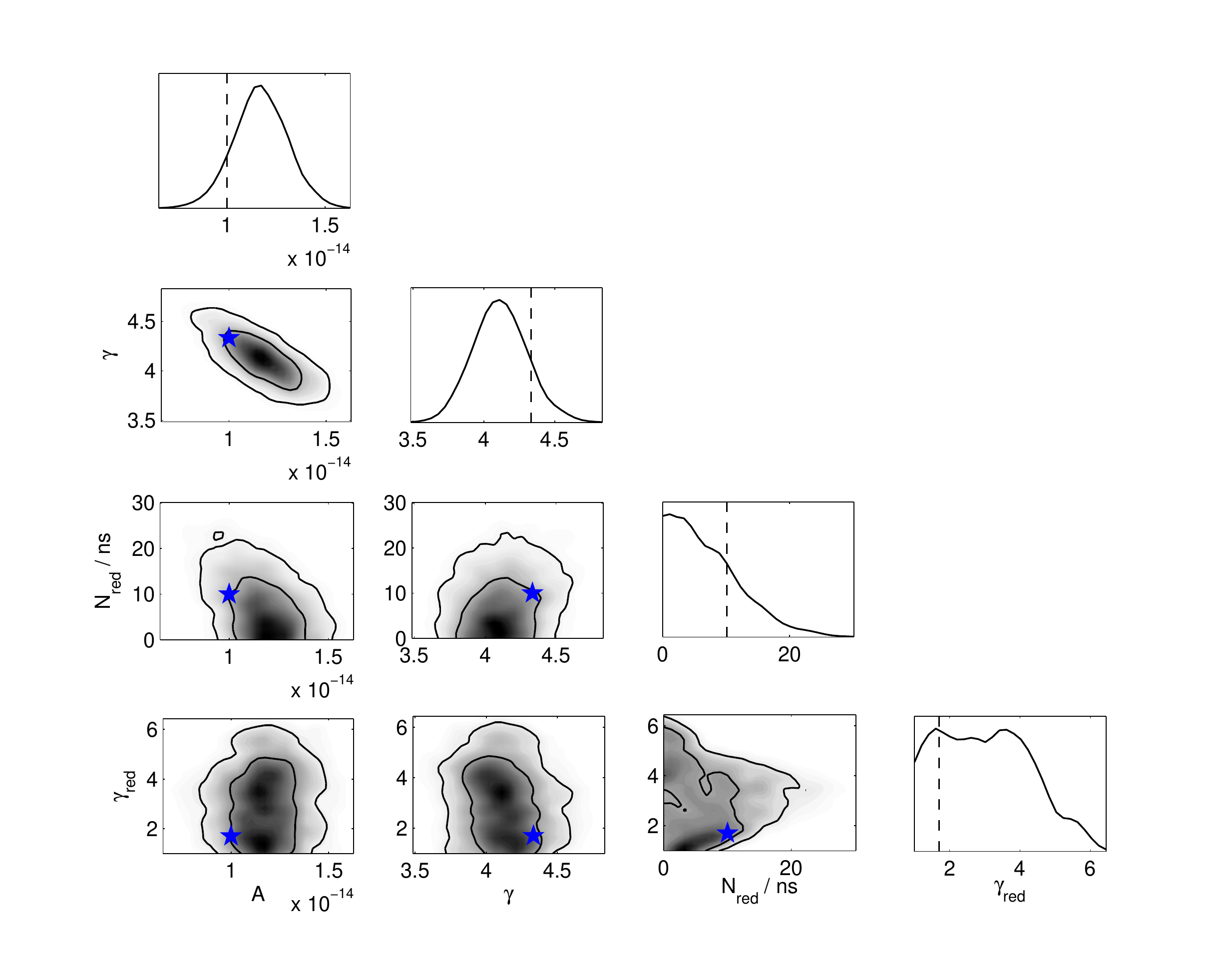}
  \caption{\label{fig:open3}The $1$D and $2$D marginalised posterior distributions of \textsc{Open}3. The stars and dashed lines show the injected values of parameters.}
\end{figure*}

This dataset had all of the complexity of \textsc{Open}2 with the addition of red timing-noise. This red-noise had a common amplitude and index for all pulsars, but was uncorrelated between different pulsars. The timing-models of all pulsars converged after a single \textsc{Tempo}2 fitting procedure.

From Table \ref{tab:all-open-models} it is clear that the evidence decisively favours the presence of a GWB. The low-frequency behaviour in the residuals was dominated by the GWB, such that in the model with a GWB and red timing-noise the red-noise properties remained largely unconstrained. The model with only a GWB gave slightly better evidence and the GWB parameter constraints were consistent with the GWB+RN model. We present results for the GWB+RN model since we know that that this is the true description of the dataset, but in a blind analysis we would not have inferred the presence of red-noise. The properties of the GWB are $A=(1.17\pm 0.14)\times10^{-14}$ and $\gamma=4.1\pm 0.2$, which are consistent with injected values. Figure \ref{fig:open3} shows the $1$D and $2$D marginalised posterior distributions for all parameters. The null model, where only TOA-errors are used to fit the timing-residuals, provides much greater evidence in this case than in \textsc{Open}1/\textsc{Open}2 since the injected GWB amplitude is five times smaller than in the previous two datasets.

\subsection{\textsc{Closed} Datasets}

\subsubsection{\textsc{Closed}1}
\begin{figure}
  \incgraph{0}{0.5}{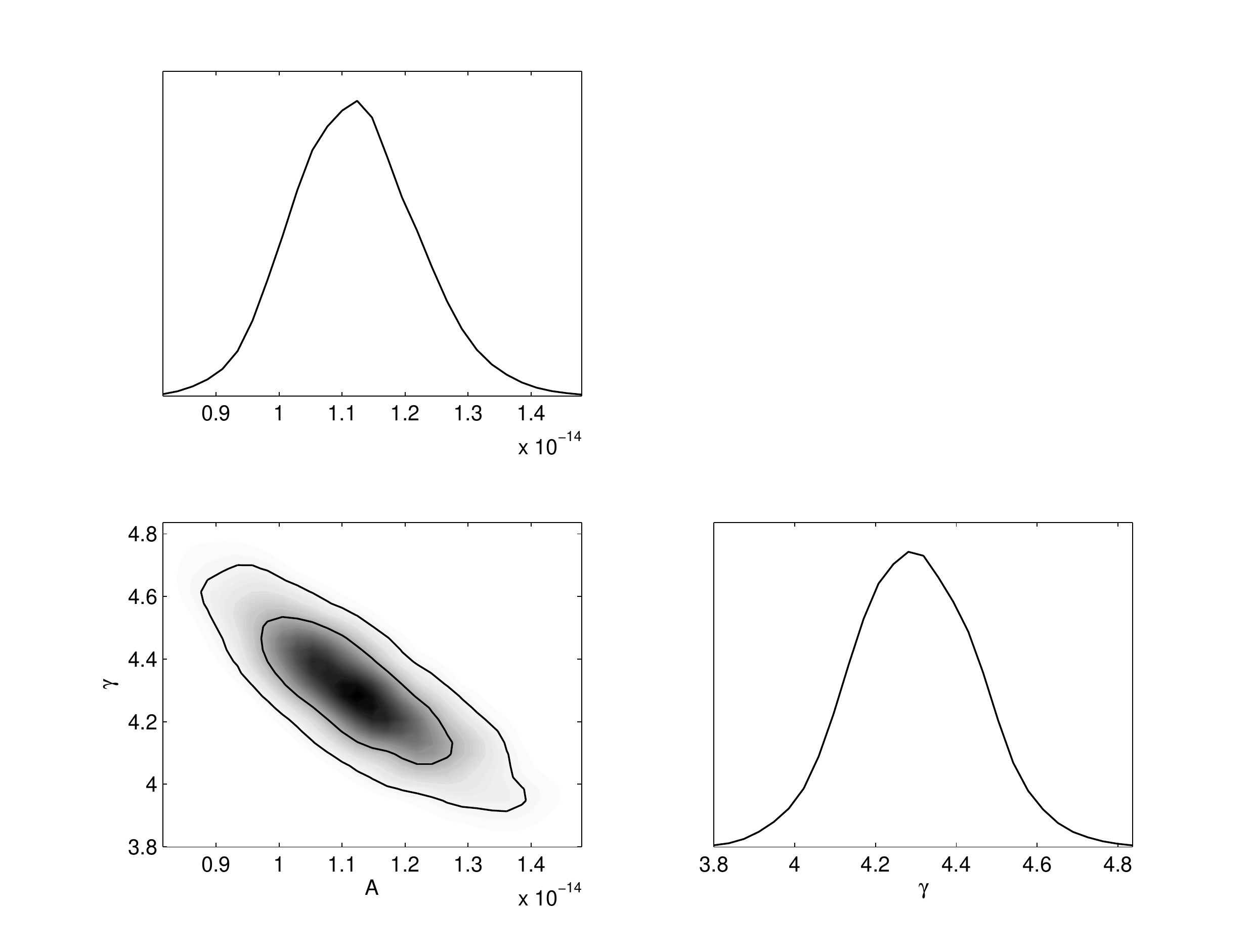}
  \caption{\label{fig:closed1}The $1$D and $2$D marginalised posterior distributions of \textsc{Closed}1.}
\end{figure}

\begin{table}\scriptsize
\caption{\label{tab:all-closed-models}Models tested for all \textsc{Closed} datasets. The most favoured model for each dataset is shown in bold. The acronyms correspond to: Gravitational-Wave Background (GWB), Time-Of-Arrival (TOA), White Noise (WN), Global EFAC (GEFAC), Global EQUAD (GEQUAD) and Red-Noise (RN).}
\begin{ruledtabular}
\begin{tabular}{l c}
Model & $\ln(\mathcal{Z})$ \\
\hline
\multicolumn{2}{c}{\textsc{Closed}1}\\
\hline
Null (TOA-errors only) & $51584.5$\\
{\bf GWB, TOA-errors} & $\mathbf{62028.1\pm 0.1}$\\
RN, TOA-errors & $62015.8\pm 0.1$\\
GWB, RN, TOA-errors & $62026.7\pm 0.1$\\
GWB, RN, GEFAC, GEQUAD, TOA-errors & $62022.7\pm 0.1$ \\
GWB, GEFAC, TOA-errors & $62025.2\pm 0.1$\\
GWB, Monochromatic-source, TOA-errors & $62028.3\pm 0.1$\\
\hline
\multicolumn{2}{c}{\textsc{Closed}2}\\
\hline
Null (TOA-errors only) & $-815557$\\
GWB, TOA-errors & $53807.4\pm 0.1$\\
GWB, RN, TOA-errors & $54149.9\pm 0.1$\\
GWB, GEFAC, GEQUAD, TOA-errors & $54178.3\pm 0.1$\\
GEFAC, GEQUAD, TOA-errors & $50082.4\pm 0.1$\\
GWB, GEFAC, GEQUAD, Burst-source, & $54235.0\pm 0.2$\\
TOA-errors & \\
GWB, GEFAC, GEQUAD, Monochromatic-source, & $54222.8\pm 0.2$\\
TOA-errors & \\
{\bf GWB, WN(J0437-4715)} & $\mathbf{55246.3\pm 0.1}$\\
GWB, WN(J0437-4715, J1455-3330, & $55235.2\pm 0.2$\\
\hspace{1.5cm}J1741+1351, J1909-3744) &\\
GWB, WN(J0437-4715), Monochromatic-source, & $55245.9\pm 0.1$\\
TOA-errors &\\
GWB, WN(J0437-4715), RN, TOA-errors & $55245.4\pm 0.1$\\
RN, WN(J0437-4715) & $55227.3\pm 0.1$\\
WN(J0437-4715) & $-240920$\\ 
\hline
\multicolumn{2}{c}{\textsc{Closed}3}\\
\hline
Null (TOA-errors only) & $53886.6$\\
RN, TOA-errors & $56168.9\pm 0.1$\\
{\bf GWB, RN, TOA-errors} & $\mathbf{56185.6\pm 0.1}$\\
GWB, RN, GEFAC, GEQUAD, TOA-errors & $56180.3\pm 0.1$\\
GWB, RN, Monochromatic-source, TOA-errors & $56185.6\pm 0.1$\\
\end{tabular}
\end{ruledtabular}
\end{table}

A list of the models tested and the associated $\ln\mathcal{Z}$ values are in Table \ref{tab:all-closed-models}. The timing-models of all pulsars converged after a single \textsc{Tempo}2 fitting procedure. We analysed \textsc{Closed}1 using the same method as \textsc{Open}1. Including a GWB and TOA errors provides evidence which, when compared to the null evidence, decisively proves the presence of a GWB. Comparing the evidence for a GWB to that for an uncorrelated red-noise process in each pulsar favours the correlated GWB. Red timing-noise in addition to the GWB is not strongly disfavoured, but did not improve the fit either. Adding in GEFAC and GEQUAD parameters did not improve the fit, as GEFAC was consistent with $1$ and GEQUAD was consistent with zero. The model with a GWB, GEFAC and TOA-errors implied a GEFAC of $0.98\pm 0.01$ i.e.,\ a $2\sigma$ deviation. However, this model did not significantly improve the fit. Including a monochromatic source did not improve the fit and all the parameters of the single source were unconstrained.

We conclude that \textsc{Closed}1 contains only a GWB with no red-noise or single GW sources. The properties of this GWB are $A=(1.1\pm 0.1)\times10^{-14}$ and $\gamma=4.30\pm0.15$. Figure \ref{fig:closed1} shows the $1$D and $2$D marginalised posterior distributions for these parameters.

\subsubsection{\textsc{Closed}2}
\begin{figure*}
  \incgraph{0}{1.0}{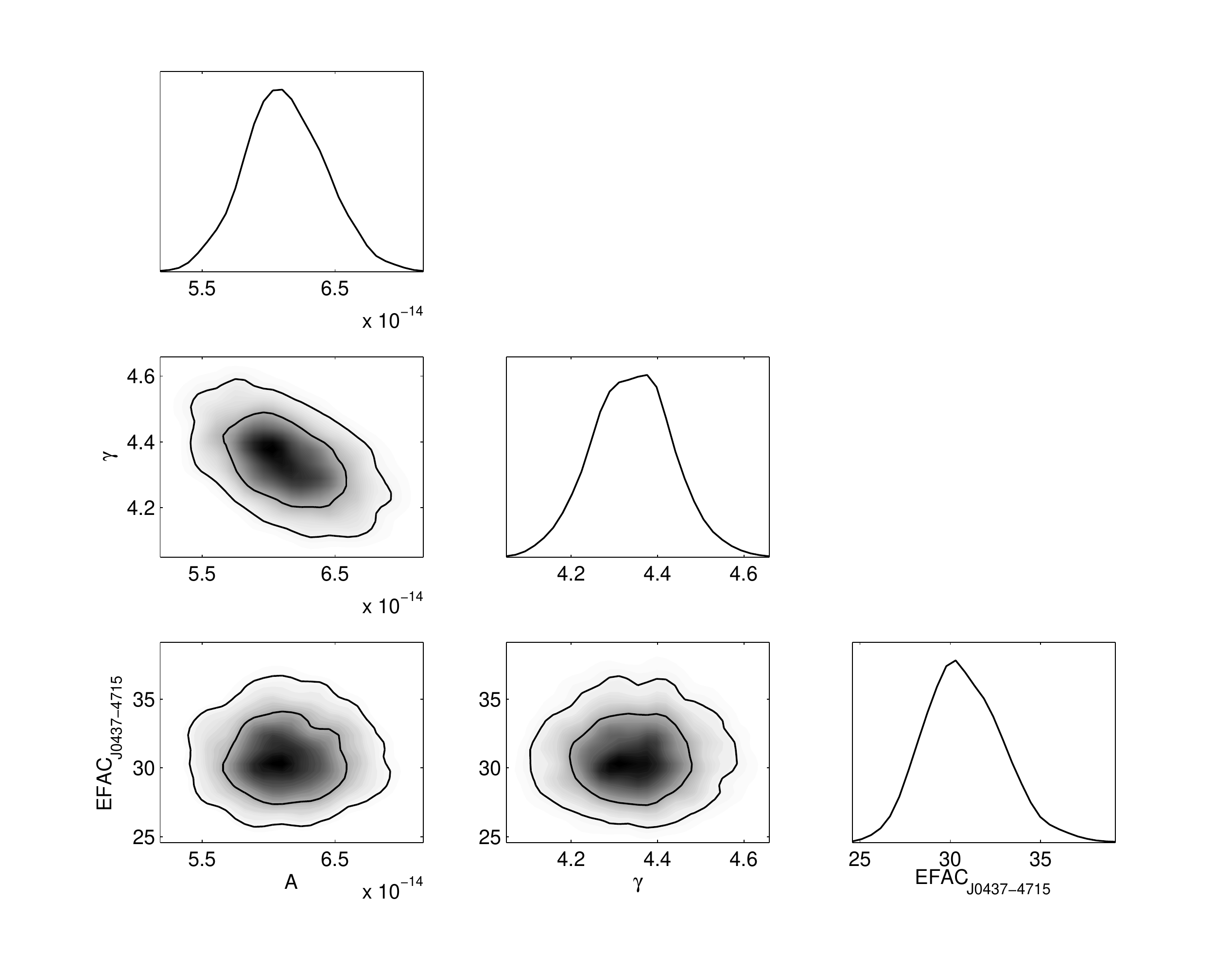}
  \caption{\label{fig:closed2}The $1$D and $2$D marginalised posterior distributions of \textsc{Closed}2.}
\end{figure*}

This dataset provided an interesting challenge. A list of the models tested and the associated $\ln\mathcal{Z}$ values are in Table \ref{tab:all-closed-models}. 

We found that assuming a model composed of a GWB and TOA-errors only did not provide a satisfactory fit. Including red timing-noise provided much higher evidence, however the index of the red-noise reached the edge of it's prior at $\gamma_{\rm red}=1$ (where the closed-form expression for the auto-covariance of a power-law PSD breaks down). A model with $\gamma_{\rm red}=0$ would effectively be a white noise model, which can also be included by the EQUAD parameter introduced earlier. The model with a GWB, GEFAC and GEQUAD provided the best evidence of these initial three models. This model implied $A=(5.9\pm 0.5)\times10^{-14}$, $\gamma=4.45\pm 0.15$, $\text{GEFAC}=0.89\pm 0.01$ and $\text{GEQUAD}=289\pm6$ ns.

We then tested for the presence of a single-source in the dataset. The evidence did appear to favour a single-source, either a monochromatic source or a GW burst. However, the recovered parameters favoured a source with a frequency several orders of magnitude above Nyquist, and a very well constrained sky-location close to the pulsar J0437-4715. As a sanity check, we removed this pulsar from the dataset and again tested for the presence of the source. In this case, all source properties were unconstrained, which is inconsistent with the presence of a single source, since the $35$ remaining widely separated pulsars should provide adequate triangulation. Furthermore, the GEFAC and GEQUAD parameters were then consistent with $1$ and $0$ respectively. Pulsar J0437-4715 has arrival-times with nominally measured timing precision of $0.03$ $\mu$s. A visual inspection of the post-fit residuals showed that such small error-bars were not sufficient to explain the high-frequency fluctuations that were present in this pulsar in addition to the red-noise induced by the GWB. We reprocessed the arrival times through several \textsc{Tempo}2 iterations, which indicated that the residuals had not converged after a single fitting iteration.

We therefore allowed the total white-noise in J0437-4715 to be a parameter to fit in our analysis. This created a very significant improvement in the evidence. Fitting for the white-noise of several other pulsars which showed ``glitchy'' behaviour did not improve the fit, so we conclude that the anomalous effects were dominated by J0437-4715. This was the only pulsar whose timing-solution did not converge after a single \textsc{Tempo}2 fitting procedure. The insufficient white-noise in this pulsar was mimicking a single, high-frequency GW-source in the pulsar's vicinity. Testing a model which fits for a GWB, the white-noise in J0437-4715 and a single-source left the source-parameters unconstrained, although the evidence was only slightly lower.

Our final results for \textsc{Closed}2 were $A=(6.1\pm 0.3)\times10^{-14}$, $\gamma=4.34\pm 0.09$. The effective EFAC on J0437-4715 was $30\pm 2$, but this was entirely due to the poor timing-model fit in this pulsar. Figure \ref{fig:closed2} shows the $1$D and $2$D marginalised posterior distributions for these parameters. When we performed a final analysis of this data, where the residuals had been cycled through \textsc{Tempo}2 multiple times to ensure convergence of the timing-model fits, we found GWB parameter constraints consistent with these results, and again no compelling evidence for a monochromatic source.

\subsubsection{\textsc{Closed}3}
\begin{figure*}
  \incgraph{0}{1.0}{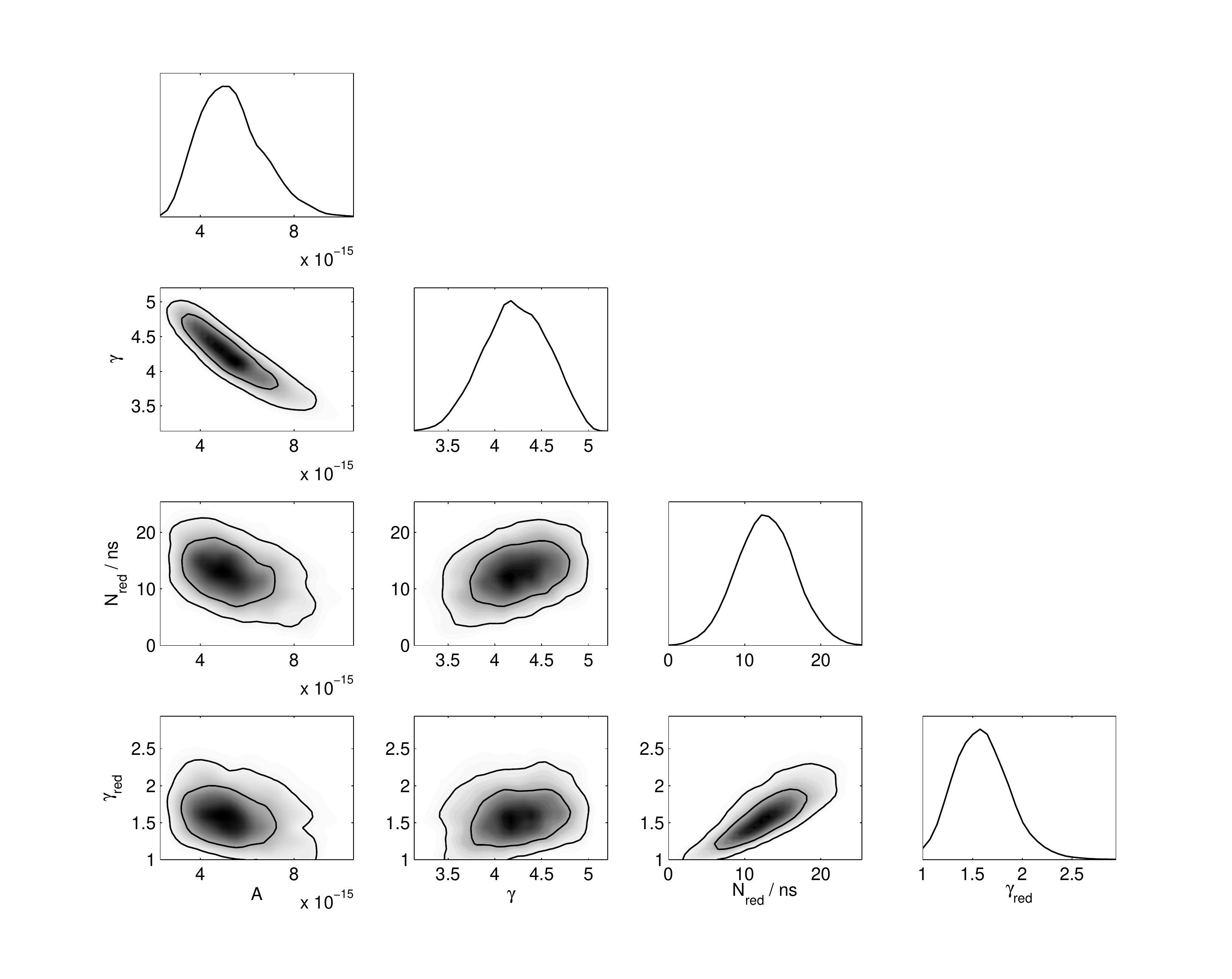}
  \caption{\label{fig:closed3}The $1$D and $2$D marginalised posterior distributions of \textsc{Closed}3.}
\end{figure*}

For the analysis of \textsc{Closed}3 we used the same model as \textsc{Open}3, including a red timing-noise component. A list of the models tested and the associated $\ln\mathcal{Z}$ values are in Table \ref{tab:all-closed-models}. The timing-models of all pulsars converged after a single \textsc{Tempo}2 fitting procedure.

The evidence decisively favours the presence of red timing-noise and a GWB. A model with GEFAC+GEQUAD gave values consistent with $1$ and $0$, respectively. A model with a monochromatic-source left the source-parameters unconstrained. We conclude that \textsc{Closed}3 contains a GWB with red-noise and no single GW sources. The properties of the GWB were $A=(5\pm 1)\times10^{-15}$ and $\gamma=4.23\pm0.35$, while the properties of the red-noise were $N_{\rm red}=(12\pm 4)$ ns and $\gamma_{\rm red}=1.5\pm0.3$. Figure \ref{fig:closed3} shows the $1$D and $2$D marginalised posterior distributions for these parameters.

\section{Acceleration by down-sampling}\label{sec:accel-down-sample}
\begin{figure*}
   \subfloat[]{\incgraph{270}{0.45}{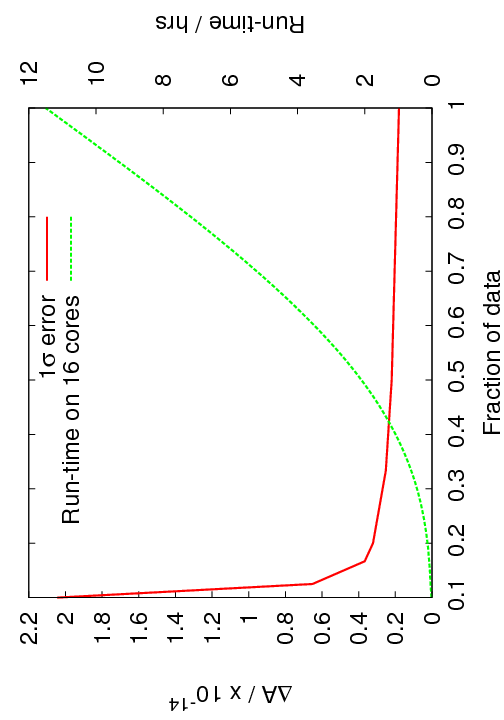}} 
   \subfloat[]{\incgraph{270}{0.45}{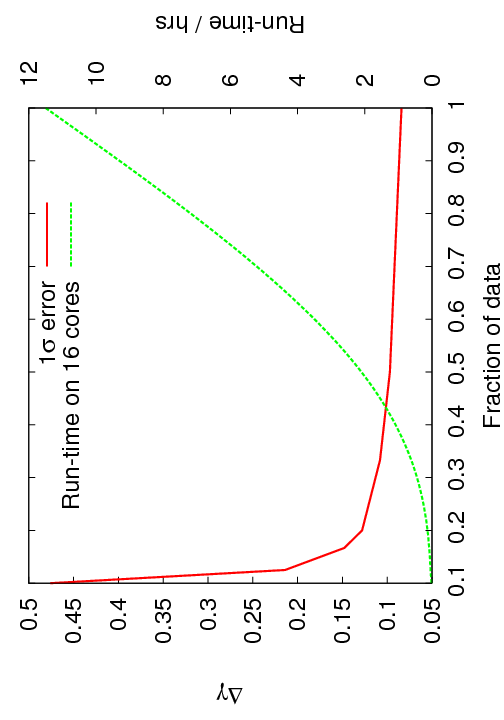}} 
   \caption{\label{fig:down-sample-error-time}\textsc{Open}1 parameter $1\sigma$ errors and run-times on a single Sandy-Bridge node with $16$ cores, for varying fractions of the data.}
 \end{figure*}
\begin{table*}
\caption{\label{tab:down-sample-error-time}\textsc{Open}1 parameter $1\sigma$ errors and run-times on a single Sandy-Bridge node with $16$ cores, for varying fractions of the data.}
\begin{ruledtabular}
\begin{tabular}{c c c c c}
Fraction of data & Cadence / weeks & \multicolumn{2}{c}{Parameter constraints} & Approx run-time\\
& & $A$ / $\times10^{-14}$ & $\gamma$ & \\
\hline
$0.1$ & $20$ & $2.04$ & $0.48$ & $1$ min\\
$0.125$ & $16$ & $0.65$ & $0.21$ & $2.5$ min\\
$0.167$ & $12$ & $0.37$ & $0.15$ & $5$ min\\
$0.2$ & $10$ & $0.32$ & $0.13$ & $8$ min\\
$0.333$ & $6$ & $0.25$ & $0.11$ & $37$ min\\
$0.5$ & $4$ & $0.22$ & $0.097$ & $2$ hr\\
$1.0$ & $2$ & $0.18$ & $0.084$ & $11.5$ hr\\
\end{tabular}
\end{ruledtabular}
\end{table*}

\begin{table}
\caption{\label{tab:mpi-multinest-speedup}\textsc{Open}1 \textsc{MultiNest} run-times for the full dataset with varying numbers of computer cores.}
\begin{ruledtabular}
\begin{tabular}{c c}
 Cores & \textsc{MultiNest} run-time\\
\hline
$16$ & $11$ hr $30$ min\\
$160$ & $2$ hr $30$ min\\
$768$ & $0$ hr $45$ min\\
\end{tabular}
\end{ruledtabular}
\end{table}

\begin{figure*}
   \subfloat[]{\incgraph{0}{0.3}{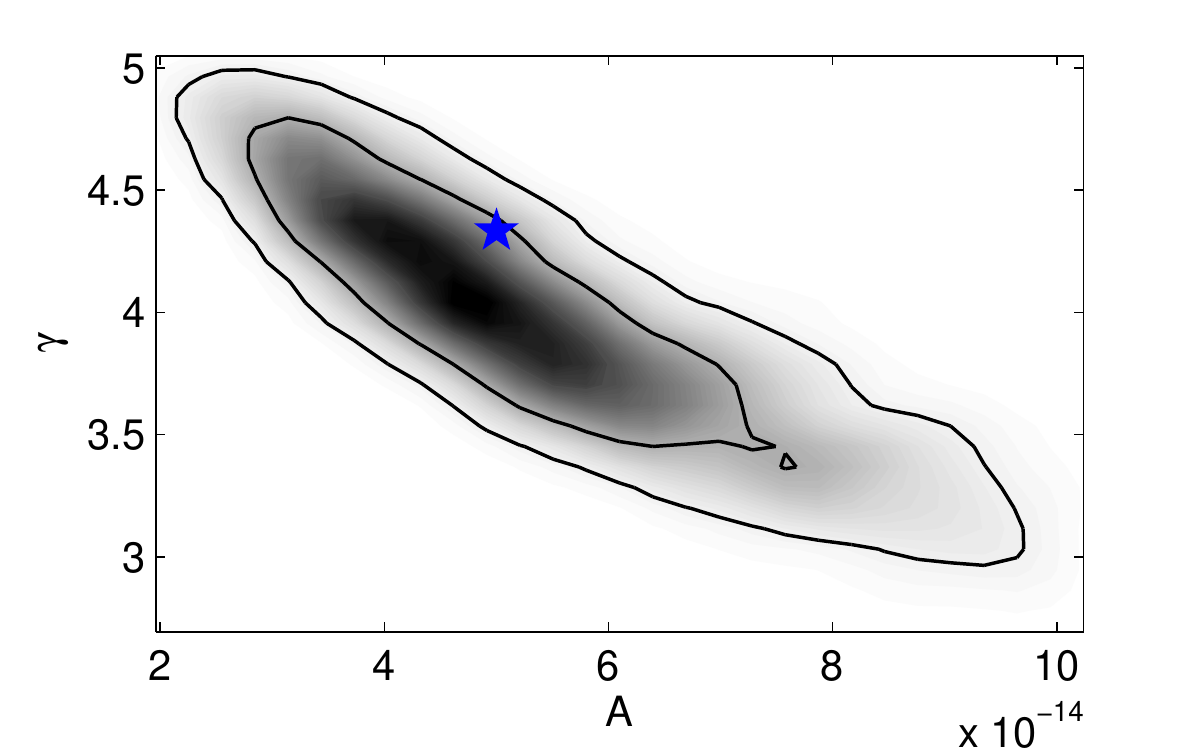}} 
   \subfloat[]{\incgraph{0}{0.3}{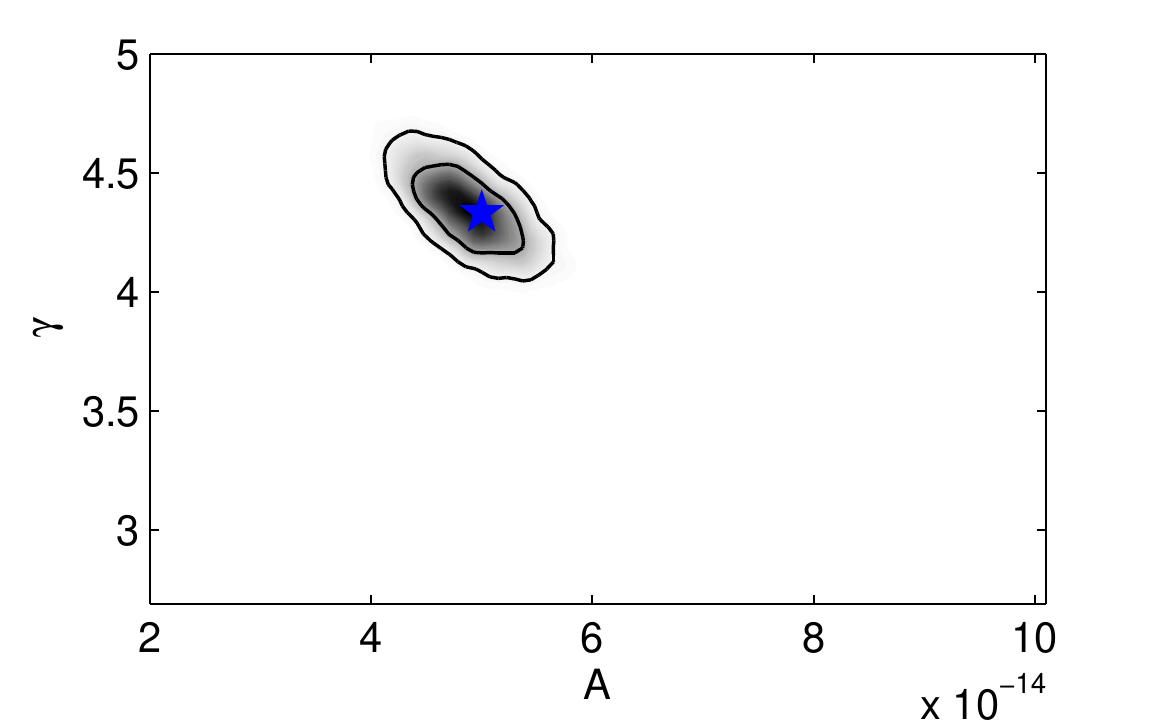}} 
   \subfloat[]{\incgraph{0}{0.3}{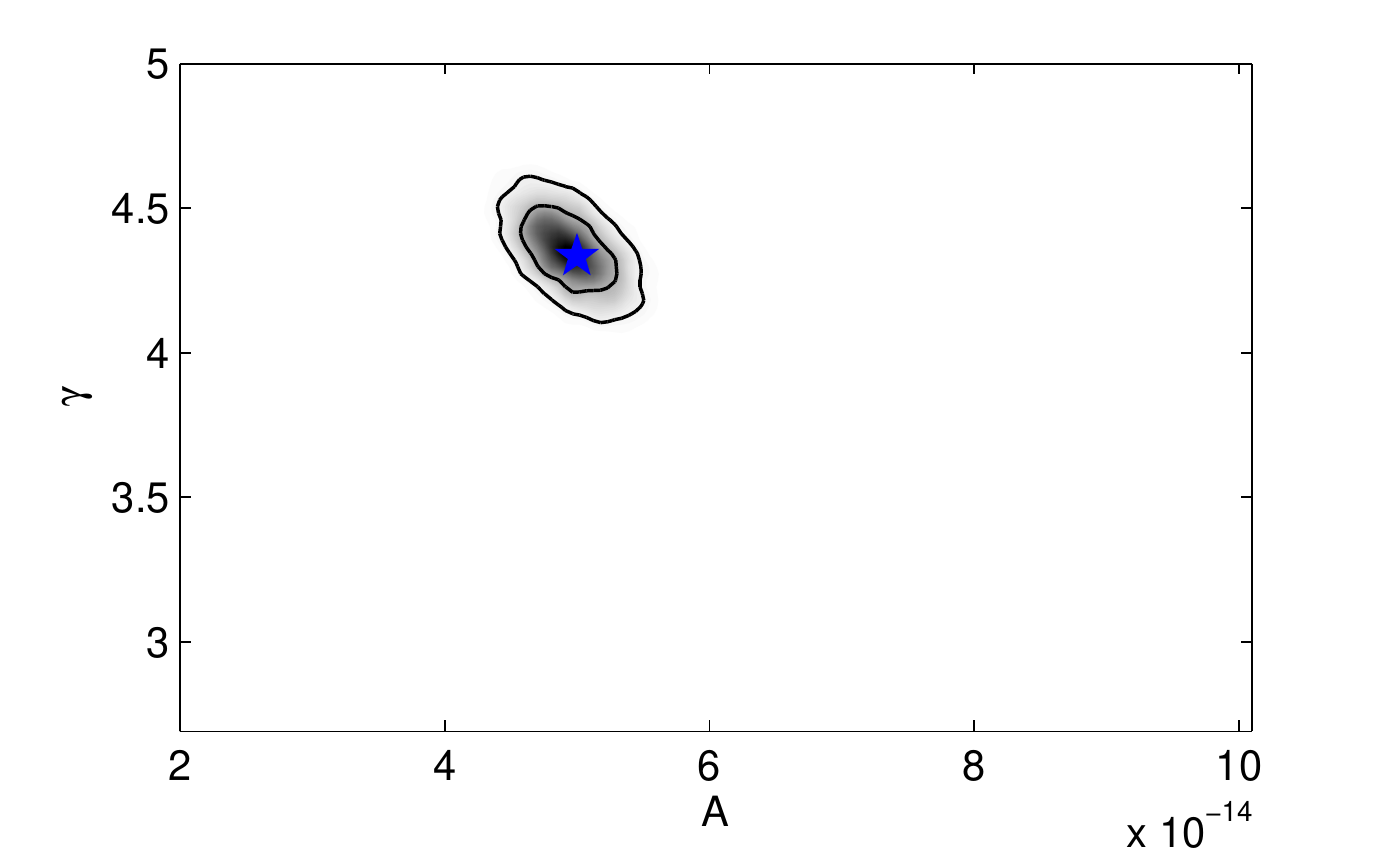}}
   \caption{\label{fig:down-sample-2d-plots}The $2$D posterior distributions with $68\%$ and $95\%$ confidence intervals for batches of data with $10\%$, $20\%$ and $50\%$ of the total \textsc{Open}1 dataset. Stars show the injected parameters.}
 \end{figure*}

The algorithm used in this work is computationally expensive and two recent proposals have been made of approaches to PTA data analysis that are faster. \citet{van-haasteren-compression-2012} used a high-fidelity data-compression technique, which utilises an interpolation scheme of the compressed covariance matrix elements to constrain $A$ and $\gamma$. The Fisher information is used to determine how much of the data might be redundant. \citet{lentati-spectrum-2012} proposed a method that avoids the dense-matrix multiplications and inversions altogether by rephrasing the likelihood in terms of matrix-vector operations and banded-matrix inversions. This is achieved by modelling the effect of the GWB on the timing residuals directly in the time-domain by a small number of independent Fourier components, effectively eliminating the off-diagonal components with $i\neq j$ in the GW correlation matrix, and avoiding an a-priori prescription for the GWB spectrum. 

Using a reduced number of frequency components is analogous to down-sampling the data in the time domain, so we investigated a very simple down-sampling of the data in \textsc{Open}1 to determine if reasonable parameter constraints could be derived when using a subset of the data. We would expect that a GWB induces a low-frequency variation of the timing-residuals, such that short cadence observations are redundant.

Table \ref{tab:down-sample-error-time} and Fig.\ \ref{fig:down-sample-error-time} show the parameter constraints obtained using down-sampled datasets comprising different fractions of the total data. The downsampling is accomplished by increasing the length of time between observations, by selecting every $n$'th time sample. The timing-model is first derived using the entire dataset then the subset of timing residuals to be used in the analysis are selected. The appropriate $G$ matrix is derived by keeping only the corresponding rows of the design matrix and deleting all other rows.

We see that, at least for the case of \textsc{Open}1, using only $20\%$ of the full \textsc{Open}1 dataset (corresponding to a cadence of $10$ weeks) is sufficient to achieve parameter constraints which are comparable to constraints from the full dataset, in approximately $\sim1\%$ of the time. The posterior widths $\Delta A$ and $\Delta\gamma$ are respectively $\sim 80\%$ and $\sim 50\%$ wider than those obtained from an analysis of the full dataset. If each $20\%$ subset of data contained the same information as the full dataset, we would expect the relative decrease in performance going from $100\%$ to $20\%$ of the data to be $\sim \sqrt{5}$, which would mean a $\sim 125\%$ increase in posterior width. We do better than this expectation since we are discarding high frequency information, while the gravitational wave background power is predominantly at lower frequencies. We do not do much better than this, however, although the analysis is much quicker. 
The corresponding $2$D posterior distributions for batches of data with $10\%$, $20\%$ and $50\%$ of the total \textsc{Open}1 dataset are shown in Fig.\ \ref{fig:down-sample-2d-plots}.

\begin{figure*}
   \subfloat[]{\incgraph{0}{0.45}{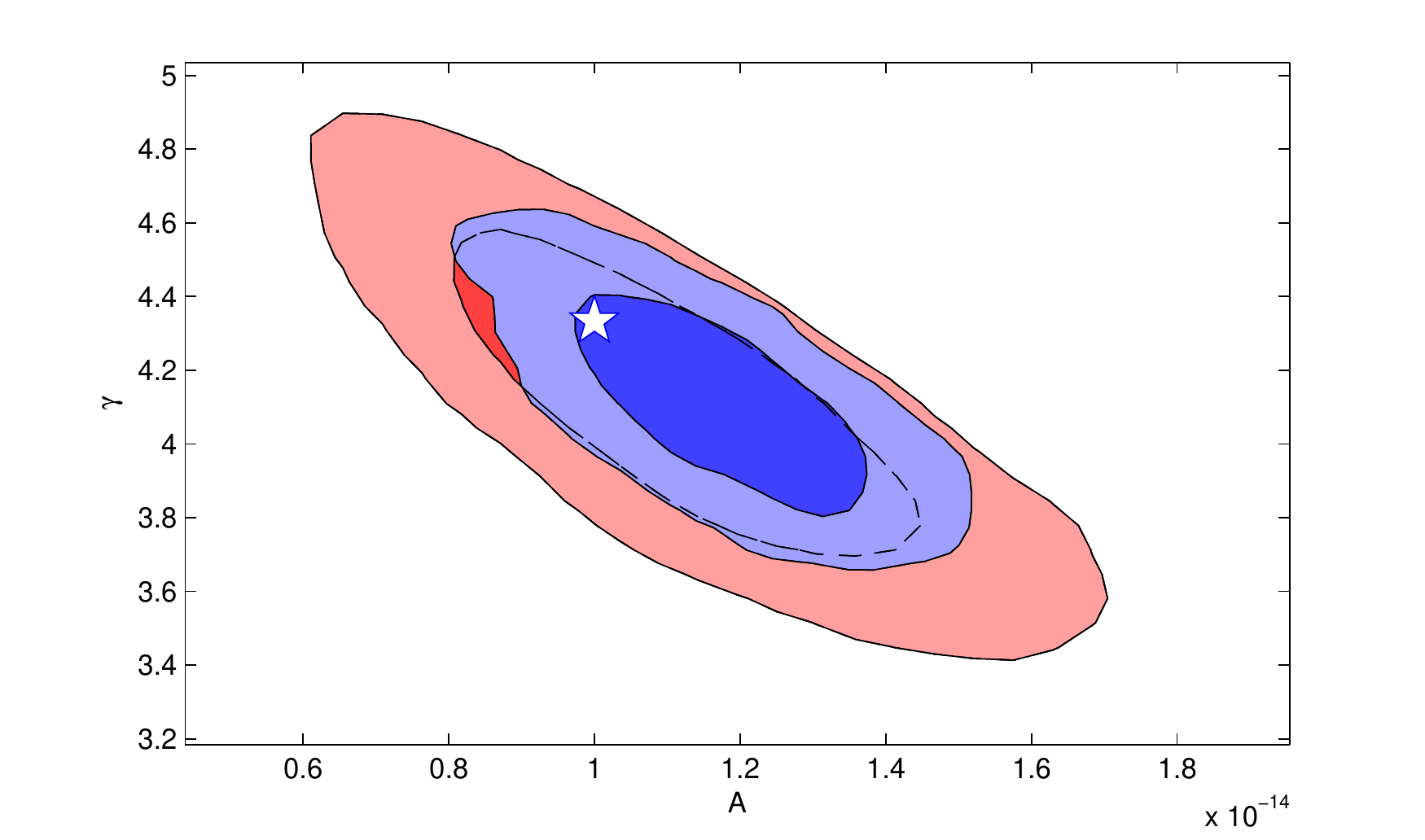}} 
   \subfloat[]{\incgraph{0}{0.45}{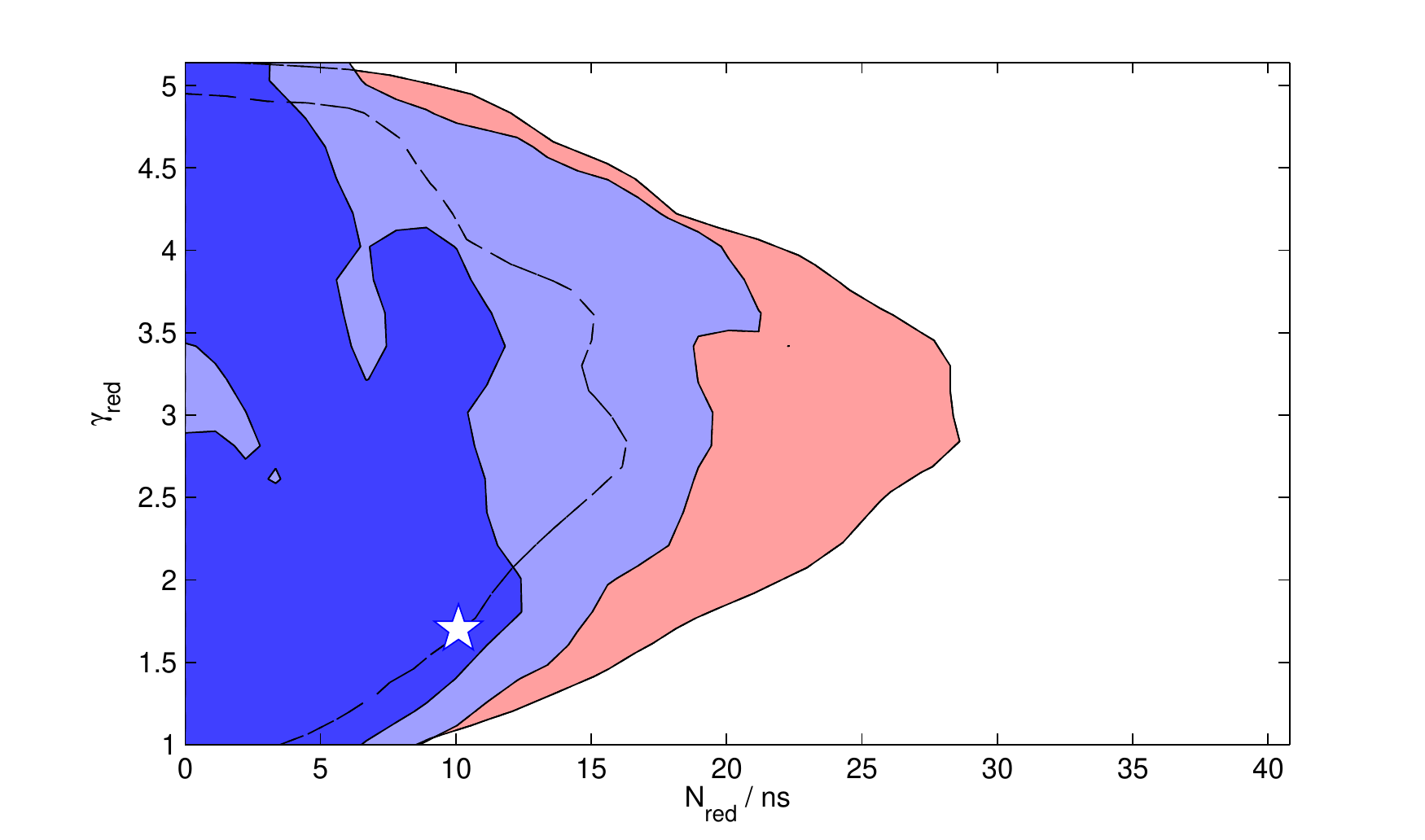}} 
   \caption{\label{fig:open3-twenty-percent}The $2$D posterior distributions with $68\%$ and $95\%$ confidence intervals for a $20\%$ batch of \textsc{Open}3 data (red) is compared to the results of a full-dataset analysis (blue). The left panel shows GWB parameters, and the right panel shows the intrinsic red-noise parameters. Stars show the injected parameters in each case.}
 \end{figure*}

For completeness, in Fig.\ \ref{fig:open3-twenty-percent} we show the result of an analysis of a $20\%$ batch of the \textsc{Open}3 data, comparing the GWB and intrinsic red-noise parameter constraints to the analysis of the complete dataset. Once again we see that, as a first-cut analysis of the data, this down-sampling technique allows us to narrow our search space when following-up with a full-dataset analysis, while also achieving comparable parameter constraints. However, much more work is required to determine the optimal way to down-sample in the more realistic case where we may have long gaps in data-taking.

We also show in Table \ref{tab:mpi-multinest-speedup} the speed-up achieved when more cores are used to exploit the built-in MPI routines in \textsc{MultiNest}. This speed-up is clearly not linear, so the appropriate compromise between computational expenditure and run-time is left at the discretion of the user.

\subsection{Stacking the down-sampled posteriors}

\begin{figure}
  \incgraph{0}{0.5}{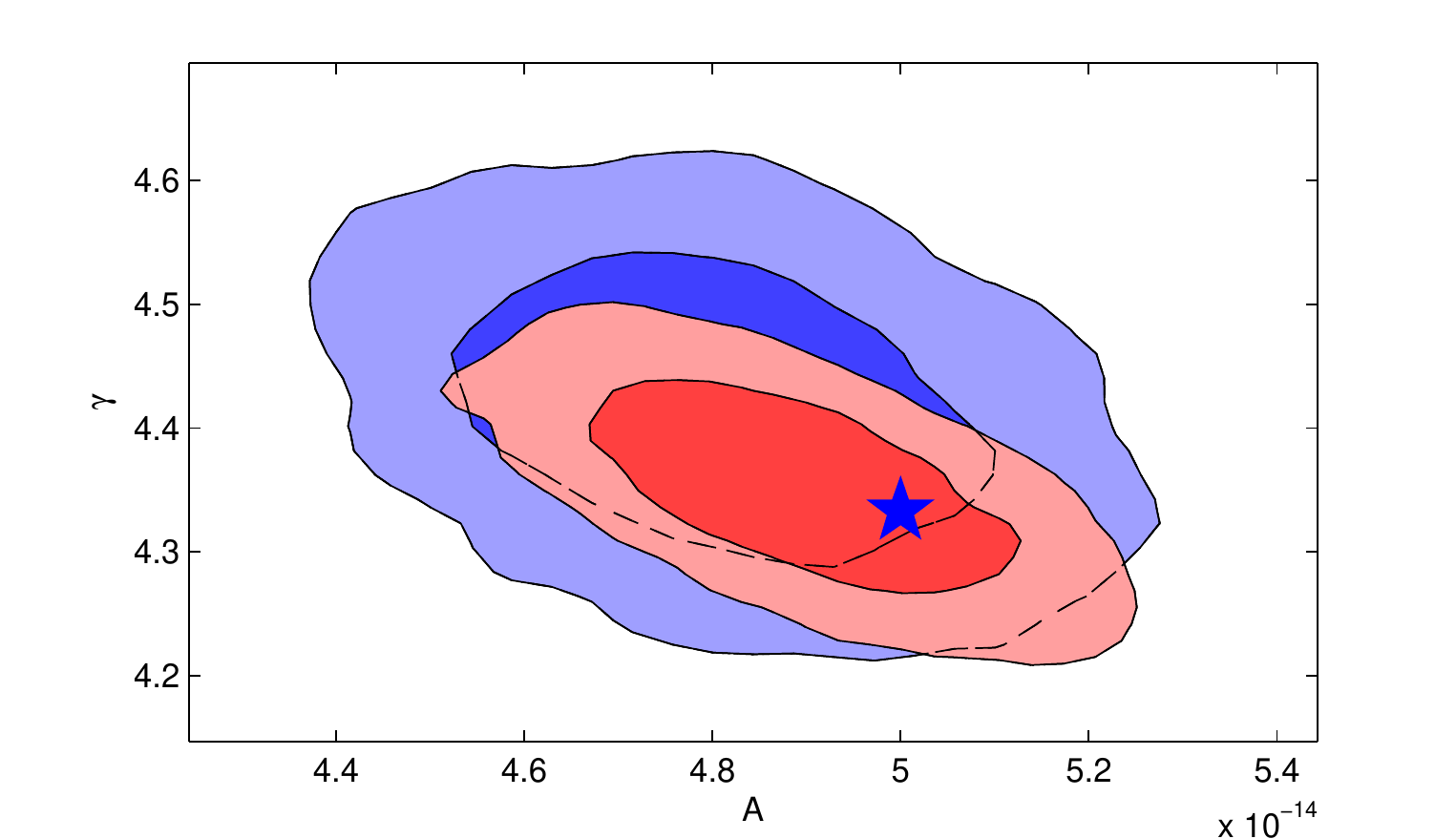}
  \caption{\label{fig:down-sample-compare}The 2D posterior distribution from a full coherent analysis of the \textsc{Open}1 dataset (blue) is compared to using the posterior distributions from the analyses of $4$ batches of $20\%$ of the \textsc{Open}1 dataset as priors on the analysis of the final batch (red). 
Credible regions are $68\%$ and $95\%$, and the injected parameters are indicated by the star.}
\end{figure}

Ideally we would like to use all of the data in our analysis, requiring a method to stack the posterior distributions from each batch of data. Unfortunately, each batch of data is not an independent sample since the post-fit residuals are the result of a \textsc{Tempo2} fit to the entire dataset. Furthermore, even without pre-fitting a timing-model to the entire dataset, if we were to combine posteriors from analyses of separate batches of raw-arrival times we would still effectively be assuming that each batch contained GWB-induced residuals drawn from a different realisation of the GWB-spectrum, which is not the case. 

Modulo these issues, it is still worthwhile to investigate the posterior widths and parameter bias that are obtained when using the na\"{i}ve approach of taking the posterior distributions obtained from analysing $(n-1)$ distinct subsets comprising $1/n$'th of the data as a prior for an analysis of the remaining $n$'th batch of data. We compared the result of using the posterior distributions from the analyses of $4$ batches of $20\%$ of the \textsc{Open1} data as priors on the final batch to the full coherent analysis of the \textsc{Open1} dataset, and this is shown in Fig.\ \ref{fig:down-sample-compare}. In this case, the stacking of the down-sampled posteriors produces a narrower final posterior, since by assuming each batch of data is an independent sample we gain a factor of $\sqrt{5}$ by combining them, while as pointed out above the individual $20\%$ posteriors are less than a factor of $\sqrt{5}$ wider than the posterior for the full dataset. The difference is not that large and the bias in the maximum a-posteriori values appears to be quite small. Most importantly, the stacked posterior remains consistent with the injected parameters in this case. For these reasons, and the fact that the analysis took $\sim 5\%$ of the run-time of the full analysis, we advocate this technique as a useful first-cut analysis of PTA data. More studies are required to understand under what circumstances the stacked posterior will no longer be consistent with the injected values.


\subsection{Recovering the angular-correlation function}

\begin{figure}
  \incgraph{270}{0.5}{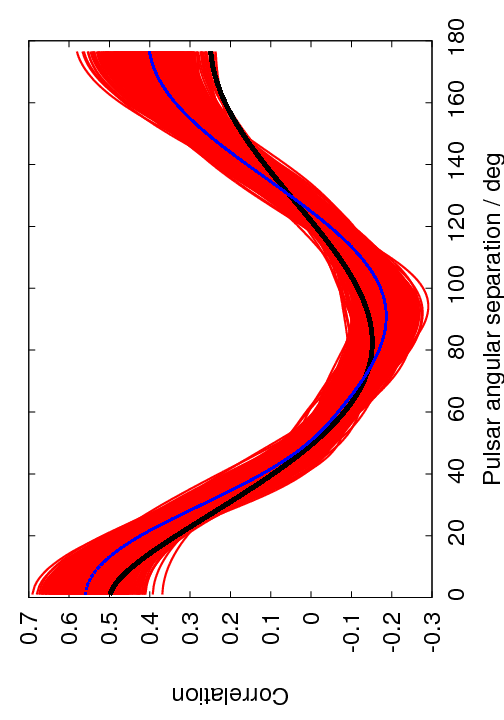}
  \caption{\label{fig:hell-downs-constrain1}The reconstructed angular-correlation of the isotropic, stochastic GWB in $20\%$ of the \textsc{Open}1 data. We parametrise the correlation at $\theta_{\rm sep}=\{1^{\circ},30^{\circ},60^{\circ},100^{\circ},140^{\circ},180^{\circ}\}$. A cubic-spline is then used to interpolate the correlation at all pulsar angular separations between $1^{\circ}$ and $180^{\circ}$, with the correlation at $\theta_{\rm sep}=0^{\circ}$ fixed at unity. The Hellings and Downs curve is shown as a black line, the max-a-posteriori correlation function is shown as a dotted line and the envelope of splines in the $95\%$ credible region of this reconstruction is shown in red.}
\end{figure}
\begin{figure}
  \incgraph{270}{0.5}{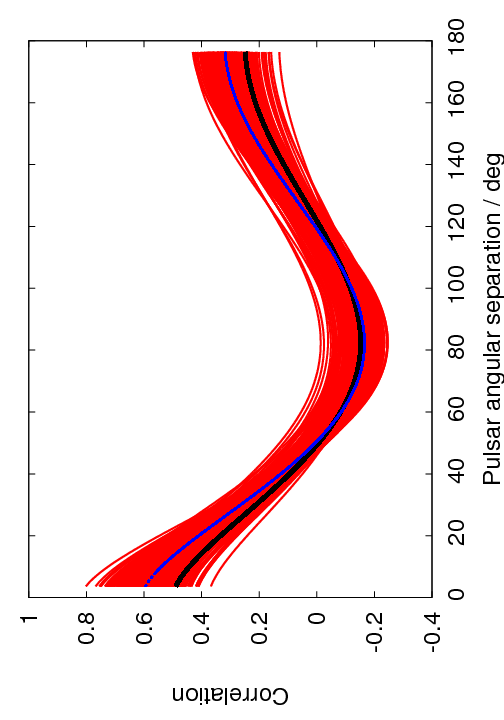}
  \caption{\label{fig:hell-downs-constrain2}The reconstructed angular-correlation of the isotropic, stochastic GWB in $20\%$ of the \textsc{Open}1 data. We parametrise the angular correlation with the same functional form as the Helling and Downs curve, but vary the numerical co-efficients as model parameters. The Hellings and Downs curve is shown as a black line, the max-a-posteriori correlation function is shown as a dotted line and the envelope of correlation functions in the $95\%$ credible region of this reconstruction is shown in red.}
\end{figure}

The acceleration of the likelihood-evaluation due to down-sampling allows us to perform a novel test. If we have an isotropic, stochastic GWB with general relativity as the correct description of gravity then the angular correlation of GWB-induced residuals will be the Hellings and Downs curve. However if we have more than the usual two general relativity polarisations then the angular correlation could deviate from this curve \citep{kj-lee-2011,chamberlin2012}. We use $20\%$ of the \textsc{Open}1 dataset to reconstruct the angular correlation of the GWB-induced timing-residuals. 

We replace the angular-correlation function $\zeta_{ab}$ by,

\begin{equation}\label{eq:hd-ansatz-1}
\zeta'_{ab} = 
\begin{cases}
1, & \text{if}\quad\theta_{\rm sep}=0^{\circ},\\
c_1, & \text{if}\quad\theta_{\rm sep}=1^{\circ},\\
c_2, & \text{if}\quad\theta_{\rm sep}=30^{\circ},\\
c_3, & \text{if}\quad\theta_{\rm sep}=60^{\circ},\\
c_4, & \text{if}\quad\theta_{\rm sep}=100^{\circ},\\
c_5, & \text{if}\quad\theta_{\rm sep}=140^{\circ},\\
c_6, & \text{if}\quad\theta_{\rm sep}=180^{\circ},
\end{cases}
\end{equation}
where $c_1,c_2,c_3,c_4,c_5,c_6$ are allowed to vary as model parameters, with prior range $\in[-0.5,1.0]$. A cubic-spline interpolation of the correlations between $1^{\circ}$ and $180^{\circ}$ is then used to calculate the correlation at all other pulsar angular-separations. The result of this analysis is shown in Fig.\ \ref{fig:hell-downs-constrain1}, where the reconstructed angular correlation is shown to be consistent with the Hellings and Downs curve. 

An alternative technique to probe the angular correlation is to assume the correlation has the same functional form as the Hellings and Downs curve, but with different numerical coefficients,
\begin{equation}\label{eq:hd-ansatz-2}
\zeta'_{ab} = p_0x\ln(x) + p_1x + p_2 + (1-p_2)\delta_{ab},
\end{equation}
where $x=(1-\cos\theta_{\rm sep})/2$ and $p_0,p_1,p_2$ are varied as model parameters, with prior ranges chosen to be symmetric around the true Hellings and Downs values. The numerical coefficients for the Hellings and Downs curve are $p_0=1.5$, $p_1=-0.25$ and $p_2=0.5$, so we chose parameter prior ranges of $p_0\in[0.0,3.0]$, $p_1 \in[-0.5,0.0]$, $p_2\in[0.0,1.0]$ respectively. However, we verified that our results were not dependent on the exact width of these priors by checking the posterior distributions was not influenced by the prior restrictions. The result of this analysis is shown in Fig.\ \ref{fig:hell-downs-constrain2}, where the reconstructed angular correlation is again shown to be consistent with the Hellings and Downs curve. 

Using the ansatzes in Eq.\ (\ref{eq:hd-ansatz-1}) and Eq.\ (\ref{eq:hd-ansatz-2}) we have reconstructed the angular-correlation of the GWB without massively expanding the dimensionality of our parameter space, and the fact that we have demanded smooth variation of the angular-correlation reduces the susceptibility of our constraints to intrinsic pulsar noise-processes. The technique which exploits the ansatz in Eq.\ (\ref{eq:hd-ansatz-1}) is more general, where we have performed a model-independent reconstruction of the angular-correlation.

This is a proof-of-principle 
that illustrates how a Bayesian framework 
can be used to infer that any GWB present in PTA data is correlated with the distinctive angular signature expected in general relativity. Further work is required to quantify how the precision of the correlation reconstruction depends on the quality of the data, the number of pulsars etc. and to explore its sensitivity to non-GR polarisation states and anisotropy in the background.

\subsection{Model-independent probes of the GWB spectrum}

\begin{figure}
  \incgraph{270}{0.5}{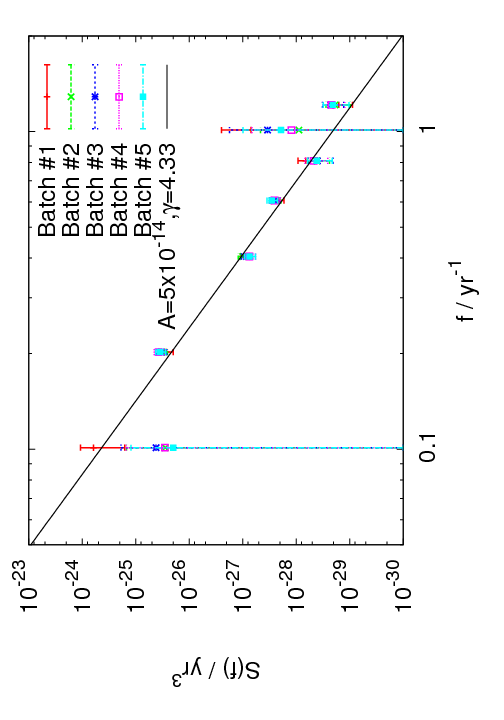}
  \caption{\label{fig:open1-spec-recover}The recovered spectrum of the GWB from the analysis of $20\%$ batches of the \textsc{Open1} dataset. We parameterise $S(f)$ at $fT=\{0.5,1.0,2.0,3.0,4.0,5.0,6.0\}$, where $T=5$ yr. We compute the autocovariance in the time-series by performing a simple trapezoidal integration over $S(f)\cos(f\tau_{ij})$, where $\tau = 2\pi|t_{i}-t_{j}|$. As expected, we have very little sensitivity to frequencies below $1/T$, however the recovered spectrum is consistent with the injected power-law spectrum.}
\end{figure}

Finally, rather than assume that the GWB spectrum is described by a power-law, we could parameterise the power-spectral density, $S(f)$, at certain frequencies to probe the spectrum in a model-independent way \citep{lentati-spectrum-2012}. One would expect this approach to fail and simply return the characteristic $\propto f^{-2}$ spectrum associated with spectral leakage that results from windowing a finite time-series, since the underlying background spectrum is steeper than $f^{-2}$. However, this model-independent technique was shown to be successful in~\citep{lentati-spectrum-2012} and we believe that spectral-leakage is avoided since the model fitting by \textsc{Tempo2} removes the low-frequency power in the time-series which would dominate the leakage signature. Meanwhile, the information about the background is preserved through our knowledge of the design-matrix, $M$, which encodes information about the fitting procedure.

Using batches of $20\%$ of the \textsc{Open1} dataset, we repeated the stochastic background search but replaced the closed-form expression for the autocovariance of a time-series induced by a power-law background by a simple trapezoidal integration over $S(f)\cos(f\tau_{ij})$ (exploiting the Wiener-Khinchin theorem),  where $\tau = 2\pi|t_{i}-t_{j}|$ and $S(f)$ is parametrised at certain frequencies. Since we are investigating a steep, red spectrum we parameterise $S(f)$ at $fT=\{0.5,1.0,2.0,3.0,4.0,5.0,6.0\}$, where $T$ is the observation span. The results are shown in Fig.\ \ref{fig:open1-spec-recover}, where we see that, as expected, we have very little sensitivity to frequencies below $1/T$, and the recovered spectrum is consistent with the injected power-law spectrum.

\section{Conclusions}\label{sec:conclusions}

We have used a Bayesian time-domain method to compute solutions to the first International Pulsar Timing Array data challenge, providing constraints on the properties of the injected isotropic, stochastic gravitational-wave background. The posterior probability distribution of the stochastic background parameters is analytically marginalised over all deterministic pulsar timing-model parameters. We tested this algorithm on the \textsc{Open} datasets, successfully recovering the values of the injected parameters.

We made use of the \textsc{MultiNest} algorithm to calculate the Bayesian evidence value for various models and to provide posterior parameter distributions. Posterior PDFs were also obtained using an adaptive MCMC code to provide a cross-check of the results. Various models were tested for each dataset, including null models where only TOA error-bars were present. The computed evidence values were then used for model-selection, determining which collection of astrophysical sources provided the best explanation for the observed pulsar TOA deviations.

The results for the \textsc{Closed} datasets were as follows. The evidence for \textsc{Closed}1 favoured a gravitational-wave background with strain amplitude at $f=1\text{ yr}^{-1}$, $A$, of $(1.1\pm 0.1)\times10^{-14}$, spectral-index $\gamma=4.30\pm0.15$ and no compelling evidence for uncorrelated red timing-noise or single-sources. The evidence for \textsc{Closed}2 favoured a gravitational-wave background with $A=(6.1\pm 0.3)\times10^{-14}$, $\gamma=4.34\pm 0.09$, and, again, no compelling evidence for red timing-noise or single-sources. Finally, the evidence for \textsc{Closed}3 favoured the presence of red timing-noise and a gravitational-wave background, with no single-sources. The properties of the background were $A=(5\pm 1)\times10^{-15}$ and $\gamma=4.23\pm0.35$, while the properties of the red-noise were $N_{\rm red}=(12\pm 4)$ ns and $\gamma_{\rm red}=1.5\pm0.3$. The adaptive MCMC sampling procedure provided consistent results in all cases.

We found it was necessary to cycle the output-residuals through the \textsc{Tempo}2 software package (which performs a weighted least-squares fit of a deterministic timing-model to the pulse arrival-times) several times for some pulsars. Timing-model fits should be inspected to ensure convergence has been attained, otherwise a poor fit in even a single pulsar can severely bias any parameter constraints.

Following our analysis of the full challenge datasets, we investigated the effect that down-sampling of the datasets has on parameter constraints and an the analysis run-time. This was achieved by only selecting every $n$th residual in the data set and design matrix. Given that the bottle-neck steps of the likelihood evaluation are $\mathcal{O}(n^3)$ matrix-matrix operations, down-sampling provided a significant speed-up in the algorithm. When tested on the \textsc{Open}1 data, we found that using $20\%$ of the full data could achieve comparable parameter constraints to the full dataset, but in only $\sim1\%$ of the time. Stacking the posterior distributions of $(n-1)$ batches of data as priors for the analysis of the remaining $n$th batch of data resulted in a small bias in the posterior width, but this was not severe and the posterior remained consistent with the injected parameters.

As a final test, we used $20\%$ of the \textsc{Open}1 data to investigate whether we could constrain the angular correlation of the GWB-induced timing-residuals. If we have an isotropic, stochastic GWB, and general relativity is the correct description of gravity, then this angular correlation should be described by the Hellings and Downs curve, Eq.~(\ref{eq:hell-down}). We parametrised the correlation at various pulsar angular-separations, employing a cubic spline interpolation to compute the correlations at all other separations. Doing so we found that the reconstructed angular correlation was not consistent with zero or full correlation, but rather showed a distinctive angular signature which was consistent with the Hellings and Downs curve. This was the first test of a novel procedure, and significantly more work is required to fully understand what can be accomplished in practice. We intend to further extend this line of inquiry, given that the determination of the angular correlation of GWB-induced residuals is key to understanding the effects of finiteness in the background \citep{ravi-2012} and possible modifications to general relativity~\citep{kj-lee-2011,chamberlin2012}.

In our future work we will explore the optimal formalism to allow both background and single-source extraction, as well as the effect that finiteness of the background has on inferred background parameter constraints and the signature of the angular correlation. This first IPTA data challenge will be followed up by more complex and realistic challenges designed to push current techniques to their limit and encourage more optimal methods to be developed. We plan to analyse these future IPTA mock datasets, and real IPTA data, employing similar techniques as used here.

Inferring the presence and properties of both a gravitational-wave background and individual gravitational-wave sources is a key aim of pulsar timing array projects. Data challenges such as this first IPTA challenge allows established members of the field and newcomers to test a variety of different approaches so that optimal techniques can be determined. By the end of the 2020's we are likely to have precision instrumental coverage over a wide range of GW frequencies, in the form of possible follow-ups to the advanced ground-based interferometers, the commissioning of a space-based interferometer, and the full operation of the SKA \citep{ska-site}, which will be the most sophisticated radio telescope ever built. These complementary instruments will all contribute to the goal of gravitational-waves becoming a precision astronomical tool.

\begin{acknowledgments}
S.R.T and L.L. are supported by the STFC. J.R.G is supported by the Royal Society. We thank Rutger van Haasteren for advice and discussions regarding the Bayesian framework used in this paper. We acknowledge the IPTA data challenge committee for their work in establishing the first IPTA data challenge. This work was performed using the Darwin Supercomputer of the University of Cambridge High Performance Computing Service (http://www.hpc.cam.ac.uk/), provided by Dell Inc. using Strategic Research Infrastructure Funding from the Higher Education Funding Council for England.
\end{acknowledgments}

\appendix 

\section{Single GW sources}\label{sec:single-source-description}

\subsection{Monochromatic sources}
A monochromatic source in the PTA-band is likely to be an SMBHB in the very early inspiral stage. \citet{sesana-vecchio-2010} have shown that it is reasonable to ignore eccentricity, spin-effects and frequency-evolution for these systems. Hence the quadrupolar approximation for GW-emission can be used to describe the waveform.

In our analysis, we use the residual signal model of \citet{sesana-vecchio-2010}. The polarisation amplitudes for a monochromatic source of frequency $f$ are,
\begin{align}
h_+ &= \mathcal{A}\left(1+\cos^2\iota\right)\cos\left(\Phi(t)+\Phi_0\right),\nonumber\\
h_{\times} &= \mathcal{A}\cos\iota\sin\left(\Phi(t)+\Phi_0\right).
\end{align}
Hence for pulsar $\alpha$,
\begin{align}
s^{\alpha}(t) &= \mathcal{R}\left[\left(1+\cos^2\iota\right)F^{\alpha}_+\left(\sin\left(\Phi(t)+\Phi_0\right)-\sin\Phi_0\right)\right.\nonumber\\
&\quad\left.+2\cos\iota\left(\cos\left(\Phi(t)+\Phi_0\right)-\cos\Phi_0\right)\right],
\end{align}
where $\mathcal{R}=\mathcal{A}/(2\pi f)$, and,
\begin{align}
F^{\alpha}_+ &= F^{\alpha}_c\cos\left(2\psi\right) + F^{\alpha}_s\sin\left(2\psi\right),\nonumber\\
F^{\alpha}_\times &= -F^{\alpha}_c\sin\left(2\psi\right) + F^{\alpha}_s\cos\left(2\psi\right),
\end{align}
where,
\begin{align}
F^{\alpha}_c =\quad&\left\{\frac{1}{4}(\sin^2(\chi_{\alpha})-2\cos^2(\chi_{\alpha}))\sin^2\theta\right.\nonumber\\
&+\frac{1}{2}\cos(\chi_{\alpha})\sin(\chi_{\alpha})\sin(2\theta)\cos(\phi-\gamma_{\alpha})\nonumber\\
&\left.-\frac{1}{4}(1+\cos^2\theta)\sin^2(\chi_{\alpha})\cos(2\phi-2\gamma_{\alpha})\right\}\frac{1}{1+\hat{n}^{\alpha}\cdot\hat{k}},\nonumber
\end{align}
\begin{align}
F^{\alpha}_s =\quad&\left\{-\cos(\chi_{\alpha})\sin(\chi_{\alpha})\sin(\theta)\sin(\phi-\gamma_{\alpha})\right.\nonumber\\
&\left.-\frac{1}{2}\sin^2(\chi_{\alpha})\cos(\theta)\sin(2\gamma_{\alpha}-2\phi)\right\}\frac{1}{1+\hat{n}^{\alpha}\cdot\hat{k}},
\end{align}
\jon{where}{and} $\{\gamma_{\alpha},\chi_{\alpha}\}$ and $\{\phi,\theta\}$ are the sky-locations of the pulsar and GW-source in spherical polar co-ordinates, respectively. The azimuthal angle in spherical polars is equivalent to right ascension, while the polar angle is related to declination by $\theta = \left(\pi/2 - \text{dec}\right)$. Further details of the polarisation basis formalism can be found in \citep{sesana-vecchio-2010,pitkin-2012}.

\subsection{Burst sources}
Various types of source could generate short-lived GW ``bursts'' in the PTA band, including cosmic strings, where cusps (sections of the string travelling at close to the speed of light) emit highly beamed radiation. Waveform models of these cusp bursts exist, and take into account the spectrum of radiation when viewed slightly off the emission axis \citep[][and references therein]{bursts-multinest-2010}. Bursts of gravitational radiation may also be emitted when two SMBHs pass close to one another on a highly eccentric orbit, which can occur shortly after a major galactic merger \citep{finn-lommen-2010}. However, for the purposes of identifying the presence of a burst we adopt a simple sine-Gaussian waveform model, which is sufficiently generic to provide a good fit to many different burst sources. The waveform is centred on a particular frequency, $f_0$, and exponentially suppressed at nearby frequencies. In the time-domain this has the form \citep{ligo-grb-search-2008,abadie-burst-2012},
\begin{align}
h_+(t) &= h_{+,0}\sin\left(2\pi f_0t + \Phi_0\right)\exp\left[-\frac{\left(2\pi f_0(t-t_b)\right)^2}{2Q^2}\right],\nonumber\\
h_\times(t) &= h_{\times,0}\cos\left(2\pi f_0t + \Phi_0\right)\exp\left[-\frac{\left(2\pi f_0(t-t_b)\right)^2}{2Q^2}\right].
\end{align}

As described before, the induced residuals at the Earth are given by modulating the metric perturbation by the PTA antenna response function for each polarisation, and then integrating over time. A sine-Gaussian integrated over time is qualitatively similar to a sine-Gaussian. We therefore adopt the approach of \citep{pitkin-2012} by searching for a residual-signal of sine-Gaussian form. Hence,
\begin{align}
s^{\alpha}(t) &= R_+F^{\alpha}_+\cos\left(2\pi f_0t + \Phi_0\right)\exp\left[-\frac{\left(2\pi f_0(t-t_b)\right)^2}{2Q^2}\right]\nonumber\\
&\quad+ R_\times F^{\alpha}_\times\sin\left(2\pi f_0t + \Phi_0\right)\exp\left[-\frac{\left(2\pi f_0(t-t_b)\right)^2}{2Q^2}\right],
\end{align}
where $R_+=\mathcal{R}\left(1+\cos^2\iota\right)/2$ and $R_\times=\mathcal{R}\cos\iota$.

\bibliography{pta_refs}

\end{document}